\documentclass[useAMS,usenatbib,usegraphicx]{mn2e}

\title[L-band spectroscopy of ULIRGs]
  {Unveiling the nature of Ultraluminous Infrared Galaxies with
  3-4~$\mu$m spectroscopy\thanks{Based
on observations collected at the European Southern Observatory, Chile (proposals
ESO 69.A-0643, ESO 73.B-0574).}}
\author[G. Risaliti et al.]
  {G.~Risaliti,$^{1,2}$
   R.~Maiolino,$^1$ 
   A.~Marconi,$^1$
   E.~Sani,$^3$
  S.~Berta,$^4$ 
  V.~Braito,$^5$
\newauthor % starts a new line in the
  R.~Della~Ceca,$^5$ 
  A.~Franceschini,$^4$ 
  and M.~Salvati$^1$\\
  $^1$INAF - Osservatorio Astrofisico di Arcetri, L.go E. Fermi 5,
I-50125 Firenze, Italy\\
  $^2$Harvard-Smithsonian Center for Astrophysics, 60 Garden st.
Cambridge, MA \\
  $^3$Dipartimento di Astronomia, Universit\`a di Firenze,
  L.go E. Fermi 2, I-50125 Firenze, Italy\\
  $^4$Dipartimento di Astronomia, Universit\`a di Padova, Vicolo dell'~Osservatorio 2, 
  I-35122 Padova, Italy\\
  $^5$INAF - Osservatorio Astronomico di Brera, via Brera 28,
  I-20121 Milano, Italy
}

\date{Released 2005 Xxxxx XX}

\pagerange{\pageref{firstpage}--\pageref{lastpage}} \pubyear{2002}

\def\LaTeX{L\kern-.36em\raise.3ex\hbox{a}\kern-.15em
    T\kern-.1667em\lower.7ex\hbox{E}\kern-.125emX}

\begin{document}

\label{firstpage}

\maketitle

\begin{abstract}
We present the results of L-band spectroscopical observations of local bright
Ultraluminous Infrared Galaxies (ULIRGs), performed with ISAAC at the VLT.
The excellent sensitivity of the telescope and of the instrument provided spectra of
unprecedented quality for this class of objects, which allowed a detailed study
of the AGN/starburst contribution to the
energy output, and of the composition of the circumnuclear absorber. 
We discuss the L-band spectral features of seven single sources, and the statistical
properties of a complete sample of 15 sources obtained combining our observations
with other published 3-4~$\mu$m spectra. Our main results are:\\
1. When a spectral indicator suggesting the presence of an AGN 
(low equivalent width of the 3.3~$\mu$m emission line, steep
$\lambda-f_\lambda$ spectrum, presence of an absorption feature at 3.4~$\mu$m)
is found, the AGN is {\em always} confirmed by independent analysis at other
wavelengths. Conversely, in all known AGNs {\em at least} one of the 
above indicators is present.\\
2. Two new diagnostic diagrams are proposed combining the above indicators, 
in which starbursts and AGNs are clearly and completely separated.\\
3. The above diagnostic techniques are possible with spectra of relatively low
quality, which can be obtained for several tens of ULIRGs with currently
available telescopes. This makes L-band spectroscopy the current best
tool to disentangle AGNs and starbursts contributions in ULIRGs.\\
4. The L-band properties of ULIRGs are heterogeneous. However, we
show that all the spectral differences among ULIRGs can be reproduced starting
from pure intrinsic AGN an starburst spectra and varying 
two parameters: the amount of dust extinction of the AGN component, and the
relative AGN/starburst contribution to the bolometric luminosity.\\
5. Using the above decomposition model, we show that 
AGNs in ULIRGs have a low dust-to-gas ratio and a dust extinction curve different
from Galactic.\\
6. The estimate of the presence and contribution of AGN in a complete sample
show that AGN are hosted by $\sim2/3$ of ULIRGs, but their energetic contribution
is relevant ($>30$\% of the bolometric luminosity) only in $\sim20$\% of the sample.

\end{abstract}


\section{Introduction}
Ultraluminous Infrared Galaxies (ULIRGs, $L_{IR}>10^{12}~L_\odot$) are
the most luminous sources in the local Universe. Moreover, they represent
the local counterparts of the class of high redshift objects dominating
the far-infrared (FIR) and sub-mm backgrounds (\citealt{sand96} and
references therein).
Understanding the origin of the huge infrared emission of ULIRGs is therefore
important in itself, given the relevance of this class of objects in the
low redshift Universe, and in order to understand the nature of the high redshift 
FIR sources, which are too faint to be studied in detail with current instrumentation
(even if the Spitzer Space Telescope is a major improvement in this direction, 
\citealt{werner}).

These considerations prompted a large interest in observational and
theoretical studies on this class of sources.
%, since their discovery
%in the IRAS surveys (s). 
The main well established
aspects on the physics of ULIRGs are: (a) the infrared (from a few $\mu$m to
$\sim1000~\mu$m) emission dominates the total luminosity, and 
is due to dust reprocessing of higher frequency radiation\footnote{
In this paper we will not discuss powerful radio loud quasars, which
have an infrared luminosity in the ULIRG range, mainly due to synchrotron emission
of relativistic particles in the jet, but are not dominated by infrared radiation.};
(b) ULIRGs are mainly found in interacting or highly irregular systems 
(\citealt{sand88}).

Two different physical processes are possible as the primary energy source
of ULIRGs: nuclear activity due to gas accretion onto a supermassive black hole 
and/or strong starburst activity. Several studies 
have been performed in the past 20 years, with the aim of understanding (a) which fraction
of ULIRGs host an AGN, and (b) which is the contribution of the AGN to the
bolometric luminosity. 

In the optical and near-IR, spectral classification of ULIRGs of the
IRAS 1~Jy sample (\citealt{kim98}) established that only a minor
fraction of ULIRGs have a clear signature of an AGN (\citealt{veill95},
\citealt{veill99opt}, \citealt{veill99ir}).
This fraction increases with luminosity from $\sim20$\% for objects with 
luminosity in the range $L_{IR}\sim10^{11}-10^{12}~L_\odot$ to $\sim50$\% for
objects with $L_{IR}>10^{12.3}~L_\odot$ (\citealt{veill95}).
A large fraction ($\sim40$\%) of ULIRGs in the 1~Jy sample 
have optical/near-IR spectra typical of
Low Ionization Narrow Emission Regions (LINERs). This classification
leaves the question about the starburst/AGN relative contribution open.

In the mid-IR, ISO spectra of bright ULIRGs provided a powerful diagnostics for the
study of the nature of these sources (\citealt{g98}, hereafter G98).
The presence of high ionization coronal lines in the mid-IR has been proven to be
an indicator of the presence of an AGN (G98, \citealt{lutz6240}), while
the presence of strong emission features due to polycyclic aromatic hydrocarbon (PAH)
molecules is an indication of starburst dominance (see next Section for further
details). 

In the X-rays, spectra of bright ULIRGs, obtained with {\em XMM-Newton} by
our group (\citealt{franc03}, \citealt{braito03}, \citealt{braito04}) and with {\em Chandra} 
(\citealt{ptak}) revealed the capability of hard X-ray spectroscopy to
disentangle the contributions of starbursts and AGNs.

The works briefly referred to above significantly enhanced our understanding of the
nature of the emission of ULIRGs, but still have relevant limitations, which can
be divided into two classes: \\
(a) Reliability of the diagnostic techniques. Most
of the AGN indicators fail in several cases to detect an existing AGN, if this is heavily
obscured by gas and dust. An extreme example of such limitations is the ULIRG
NGC~6240, which has no indication of the presence of an AGN at any wavelength 
with the exception of the hard X-rays,  where the AGN direct emission
escapes the high column density ($N_H>10^{24}$~cm$^{-2}$) absorber at $E>10$~keV 
(\citealt{vignati}), and the flat reflected emission, with a prominent iron
$K\alpha$ emission line with equivalent width $EW>1$~keV are observed in the 2-10~keV range.\\
%(LIST : Genzel (Mid IR), Veilleux (Opt NIR) X-rays (Franceschini, Ptak), L-band
%(Imanishi, Risaliti).\\
(b) Extension to faint sources. The most powerful indicators of the presence
of buried AGNs in ULIRGs have been proven to be the hard X-ray spectra, and ISO
mid-IR spectra. In both cases a good signal-to-noise spectrum is needed in order
to detect the AGN/starburst indicators: in the X-rays, a high S/N is required
to disentangle the hard power law due to the AGN emission from the thermal component
due to the starburst; in the mid-IR a high S/N is needed to have an accurate
determination of the continuum and, consequently, a reliable measurement of
the equivalent width of the PAH emission features. With current infrared detectors, 
the required spectral quality can be achieved only for a small number ($\sim10-15$) of bright 
objects. A big improvement for the mid-IR analysis is expected thanks to
the Spitzer Space Telescope, which will be able to provide high quality
mid-IR spectra for several tens of ULIRGS, as first results have already confirmed
(\citealt{armus}).

An interesting new way of studying ULIRGs has been proven to be L-band ($\sim3-4~\mu$m) 
spectroscopy. Imanishi and collaborators (in particular, Imanishi \& Dudley~2000) 
discussed the physics of the emission/absorption
features in the 3-4~$\mu$m interval, and showed that low-resolution spectra
of the brightest ULIRGs, obtained at 4 meter class telescopes, provide effective
diagnostic elements in order to disentangle the AGN and starburst contributions to
the observed emission. Recently, Imanishi, Dudley \& Maloney~(2005) applied this 
method to investigate the energy source in a sample of $\sim40$ objects from
the 1~Jy sample (Kim \& Sanders~1998).

Risaliti et al.~(2003) presented a high signal-to-noise L-band spectrum of the bright ULIRG
IRAS~19254-7245 obtained with the instrument ISAAC at the Very Large Telescope (VLT) 
in Paranal, Chile. This high quality spectrum suggested that 8 meter class telescopes are able to
provide good enough spectra for (a) a detailed study of the physical properties
of bright ULIRGs, and (b) an analysis of relatively faint ULIRGs, for which other indicators
are not available with the current instrumentation. 

In this paper we present new L-band spectra of bright ULIRGs obtained with the instrument ISAAC
at the VLT in Paranal, Chile, and we perform a complete analysis of
the L-band diagnostics applied to a representative sample of ULIRGs, consisting 
of our new observations and previously available L-band spectra.
In Section 2 we review the L-band spectral features 
useful in order to investigate the energy source in ULIRGs. 
In Section 3 we present our sample. 
In Section 4 we present the new L-band spectra.
The physical analysis of each new spectrum is presented in Section~5. 
In Section~6 we present a detailed discussion of our results, and an analysis
of the different indicators in a complete flux-limited sample, obtained merging
our sample of southern sources with other spectra obtained by Imanishi and collaborators
for northern ULIRGs. In particular: (1) we present new two-dimensional diagnostic diagrams
where starburst and AGNs are completely separated (2) we discuss the physical
properties of the dusty circumnuclear medium in ULIRGs hosting AGNs, showing that
a different amount of hot dust absorption can explain the main differences among
spectra of these sources; (3) we show that the amount of hot dust can also explain the
differences among starburst-dominated ULIRGs.

Our conclusions are summarized in Section~7.
Through this paper we estimate luminosity distances using the WMAP concordance 
cosmology ($h_0=70, \Omega_m=0.3, \Omega_\lambda=0.7$, e.g. \citealt{spergel}
%%%%%%%%%%%%%%%%%%%%%%%%%%%%%%%%%%%%%%%%%%%%%%%%%%%%%%%%%%%%%%%%%%%%%%%%%%%5

\section{L-band diagnostics for ULIRGs}

Disentangling the contribution of heavily obscured AGNs and starbursts
to the luminosity of ULIRGs is intrinsically complicated, because 
of the reprocessing, in both cases, of the primary optical/UV
radiation into mid- and far-IR emission. 

However, several physical differences between the two energy sources 
can have direct consequences on the observed reprocessed radiation
in the infrared: 

\begin{itemize}
\item The X-ray emission of an AGN, much stronger than that of a starburst
(with respect to the bolometric emission) affects the gas and dust composition
around the central source. For example, (a) highly ionized atomic species, requiring
a strong ionizing X-ray continuum,
are present only around AGNs, and (b) complex hydrocarbon dust grains can be present
only around starbursts, while they cannot survive in the close vicinity of AGNs, since 
they are destroyed by the X-ray photons.
\item An active nucleus is an extremely compact radiation source with respect to
a starburst region. This implies that strong absorption effects in AGN spectra are
easily obtained through compact and dense absorbers along the line of sight.
The same effects are harder to obtain in starbursts, since the absorber should
be large enough to cover the whole emitting region.
\item The amount of high temperature (from a few 10$^2$~K to the sublimation temperature,
$T_{sub}\sim1,500$~K) dust is higher in AGNs than in starbursts. This affects the
infrared spectral energy distribution, in particular in the near- and mid-IR.
\end{itemize}

In particular, the emission in the L-band is due
to reprocessing of the primary radiation by hot dust (temperature in the range 100-1000~K), 
i.e. by material relatively close to the central source. As a consequence,
the observed L-band spectra are heavily affected by the above physical differences.

Here we briefly review the main diagnostic features in ULIRGs, most of which are
present in some of the spectra shown in Fig.~1.
%\ref{plot1}. 
These indicators have been calibrated
on objects for which the starburst/AGN classification is known from other, independent
studies. However, several cases are known for which some of these diagnostic tools fail.
We will discuss these cases, and possible improvements of the diagnostic techniques,
in Section~6.\\

{\bf 3.3$\mu$m emission feature.} A strong, broad emission feature at a rest-frame
wavelength $\lambda=3.3\mu$m is the most prominent non-continuum feature in the
spectra of most ULIRGs. 
The origin of this emission feature is believed to be the reprocessing of UV radiation
by polycyclic aromatic hydrocarbon (PAH) molecules. These complex organic molecules 
(for a review on their structure and physical properties, see \citealt{all89}) are responsible
of several broad emission features in the mid-IR wavelength band, and have
been extensively studied in ISO spectra of ULIRGs (e.g. G98).

The 3.3~$\mu$m emission feature is particularly interesting in this context because 
it is an isolated feature, which can be easily disentangled from the continuum emission.
This is not the case in many mid-IR ISO spectra, where the combination of broad
emission features due to PAHs with broad absorption features due to dust makes
the determination of the continuum level (and therefore of the equivalent
width of the lines) extremely uncertain (one notable exception is the PAH
emission feature at 6.2~$\mu$m, as discussed in \citet{fischer}).

Both theoretical and observational evidence suggests that the strength of
the 3.3$\mu$m PAH feature is an indicator of the relative contribution
of AGN and starburst to the emission of ULIRGs:\\
- Observationally, it is found that objects which are independently classified
as starburst-dominated show a 3.3~$\mu$m feature with an equivalent width 
$EW_{3.3}\sim$100~nm or higher.
This is true both at ULIRGs luminosities and for objects with luminosities
of the order of $\sim10^{11}~L_\odot$ (\citealt{moor86}, \citealt{im-du}).
Objects known to be dominated by AGN emission show a weak or absent ($EW_{3.3}<30$~nm)
3.3~$\mu$m emission feature (\citealt{im-du}). 
As we will discuss in more detail later, this is not
a ``perfect'' indicator. However, the correlation is strong, and no exception is
known at least in cases of extreme values (i.e. all known sources with $EW_{3.3}<50$~nm
or $EW_{3.3}>100$~nm are dominated by AGN or starburst, respectively).\\
- From a physical point of view, two mechanisms are expected to be effective in
decreasing the equivalent width of the 3.3$\mu$m line in AGNs: (a) the X-ray
radiation can destroy the PAH molecules which are not shielded by a column density
from a few $10^{22}$ to 10$^{24}$~cm$^{-2}$ of gas, depending on the 
geometrical and physical conditions (\citealt{voit}), and (b) the strong L-band continuum
(much stronger than in starbursts due to the presence of a larger amount of hot dust)
can dilute the 3.3$\mu$m emission from outer or gas-shielded starburst regions.\\

{\bf 3.4$\mu$m absorption feature.} A deep absorption feature is clearly present
in the spectrum of several ULIRGs. The most prominent cases are IRAS~19254-7245S
(Fig.~1, see \citealt{risa03} for details), UGC~5101 (\citealt{im5101x}) and
IRAS~08572+3915 (\citealt{im-du}), where the optical depth\footnote{All the optical
depth estimates presented in this paper are computed as the natural logarithmic ratio between
the estimated continuum flux and the measured emission at the wavelength corresponding to the 
mimimum of the absorption profile.} at 3.4~$\mu$m is
$\tau_{3.4}\sim0.8-1$. 
The origin of this feature is believed to be absorption by hydrocarbon dust grains,
in particular by vibrational stretching levels of the C-H$_2$ and C-H$_3$ groups
(\citealt{sandford} and references therein, \citealt{pend02}).
In the highest signal-to-noise detections, as the ones mentioned above,
a detailed study of the chemistry of the absorbing dust is possible
through the analysis of the substructure of the absorption profile (\citealt{risa03},
\citealt{mason}).
 
The presence of the 3.4$\mu$m absorption feature is an indicator of a strong AGN contribution
to the ULIRG emission. Again, the motivation has both a physical explanation
and an observational confirmation:\\
- Observationally, all the above mentioned ULIRGs are independently known
to host a powerful AGN. The other known cases among ULIRGs 
where this feature has been revealed (\citealt{im-du}),
even if with a smaller optical depth, are all classified as AGNs on the basis
of independent classification at other wavelengths. Furthermore, L-band spectra
of non-ULIRG obscured AGNs show strong 3.4$\mu$m absorption (\citealt{im-agn}).\\
- Physically, in order to detect a high 3.4$\mu$m absorption optical depth,
a high column density absorber in front of the primary source 
is required.
\citet{im-ice} showed that dilution and saturation effects
prevent the observed optical depth from being higher than $\sim0.2$ in
case the absorbing dust is spatially mixed with the energy sources.
Therefore, in order to achieve high values of $\tau_{3.4}$, an external
screen completely covering the radiation source is needed.
Since the absorbing dust is expected to lie in gas clouds, 
we can estimate the amount of gas expected to be associated with the
dust
assuming a Galactic dust-to-gas ratio, and the extinction curve of 
\citet{pend94}. We find  
$N_H\sim3-5\times10^{23}\tau_{3.4}$~cm$^{-2}$. The actual
value could easily be more than an order of magnitude greater, given that
the measured dust-to-gas ratio in AGNs is systematically lower than Galactic
(\citealt{almud}, \citealt{macc82}, \citealt{maio}). 
Such high column densities are easily present around AGNs ($\sim$half AGNs in
the local Universe are absorbed by column densities higher than 10$^{24}$~cm$^{-2}$,
\citealt{risa99}), while a complete
covering of a starburst region would imply a huge amount of gas:
assuming the linear
dimension of a starburst region emitting $\sim10^{12}~L_\odot$ to be of the order of 
or greater than one kiloparsec, 
the required mass of gas would be $M_{gas}>10^{11}\times (R/100~{\rm pc})^2~M_\odot$.\\

{\bf 3.1$\mu$m absorption feature.} 
A broad absorption feature at $\sim3.1~\mu$m has been detected in several ULIRGs
(\citealt{im-ice}). The origin of this feature is absorption by ice-covered
dust grains. Analogously to the 3.4$\mu$m feature, it can be shown that
optical depths $\tau_{3.1}>0.3$ are not possible unless an external screen
covering the source is present, if a fraction of 30\% of dust grains covered by ice
is assumed (as estimated for the nearby starburst M82, Imanishi \& Maloney~2003). 
Then the same argument as above suggests that
this is highly unlikely in starburst-dominated sources.
This expectation is confirmed by the discovery of broad ice absorption in AGN-dominated
ULIRGs (\citealt{im-ice}). 
However, the uncertainty on the possible profiles and widths of this absorption
features makes it less straightforward than the previous indicators, especially
in low signal-to-noise spectra. In particular,
in many cases it is not possible to distinguish in AGN-dominated objects 
whether a concave spectrum in
the 3-3.6$\mu$m region is due to a broad ice absorption feature, or to a change in
spectral slope (in a $\lambda-f_\lambda$ plane) between the near-IR region
(still dominated by the tail of the direct emission of the AGN accretion disc)  
and the mid-IR region (dominated by reprocessing of hot dust).
In order to distinguish between these scenarios, a high signal-to-noise spectrum
is required, in order to constrain the continuum slope using the 
feature-free 3.5-4~$\mu$m wavelength region. Alternatively a broader wavelength
coverage, involving K-band and/or M-band spectra are needed. 
Moreover, a different (higher) fraction of dust grains covered by ice could increase
the upper limit of $\tau_{3.1}$ in starbursts. 

In the next Section we will separately discuss each case among our spectra
where ice absorption could be present.\\

{\bf Continuum slope.} 
The observed continuum slope in the L-band spectra of ULIRGs has a wide range,
spanning from $\Gamma=\sim-2$ to $\Gamma\sim3$ when fitted with a $f_\lambda=K\lambda^\Gamma$
model.
Observations of known AGNs and starbursts suggest that positive $\Gamma$ are an
indication of AGN activity.
The physical explanation of this trend is clear:
a high $\Gamma$ implies strong dust reddening of a compact source,
and hence the presence of an AGN.
As for the other indicators, we note that observations confirm the
effectiveness of the continuum slope in extreme cases
($\Gamma>2$ are always associated to AGNs), while
objects with $\Gamma<2$ (i.~e. most of the ULIRGs observed so far) can
be either AGNs or starbursts.
We will go into further details on this point in Section~6.\\

{\bf Atomic emission lines.} Atomic emission lines are in principle powerful diagnostics
of the presence of an AGN. A broad atomic emission line, such as the $Br\alpha$
line at 4.05~$\mu$m, or a high ionization line such as the [SiIX]~$\lambda3.95~\mu$m
or the [MgVIII]~$\lambda3.028~\mu$m,
are strong indicators of the presence of an AGN. These diagnostics have been successfully
used on nearby Seyfert~2 Galaxies (\citealt{lutz-m}).

However, out of these three lines, by far the
most intense expected in the (rest frame) L-band, two ($Br\alpha$ and [SiIX]) are in all 
our objects out of the well calibrated 2.9-4.1~$\mu$m interval, due to the redshift effect,
and the third one, [MgVIII] is in the poor atmospheric transmission interval (Fig.~1).
Therefore, we will not use this diagnostics in this work.

%###################################################################

\section{The bright ULIRGs sample}

The sample chosen for L-band spectroscopic observations is the one of
G98, consisting of 15 ULIRGs in the IRAS Bright Galaxy
Sample (BGS, consisting of all the galaxies in the IRAS all sky survey 
with galactic latitude $\|b\|>5^o$ and
60~$\mu$m flux density $S_{60}>5.24$~Jy, \citealt{bgs}), 
with two variations: one source 
(IRAS~23060+0305) has been dropped because it does not fulfill the flux
requirement of the BGS, and two sources (IRAS 08572+3915 and IRAS 05189-2524)
have been added, because they have $S_{60}>5.24$~Jy (Table~1).

The luminosity criterion used to select ULIRGs is the standard one,
i.e. $L_{IR}>10^{12}~L_\odot$, where $L_{IR}$ is the total 8-1000$\mu$m luminosity,
calculated using the IR flux equation of \citep{sand96}:
\begin{equation} 
F_{IR}=1.8\times10^{-11}(13.48S_{12}+5.12S_{25}+2.58S_{60}+S_{100})
%erg~s^{-1}cm^{-2}
\end{equation}
where $F_{IR}$ is the total IR flux in units of erg~s$^{-1}$~cm$^{-2}$,
and $S_{12}, S_{25}, S_{60}, S_{100}$ are the flux densities in the four
IRAS filters, in units of Jy.

One source, NGC~6240, has a slightly lower total infrared luminosity.
However, it is included in our sample, as well as in that of G98, because
all the morphological and physical properties appear to be typical of
ULIRGs.

\begin{table*}
\centerline{\begin{tabular}{lccccccccccc}
\hline
Name & z & S$_{12}^a$ & S$_{25}^a$ & S$_{60}^a$ & S$_{100}^a$ & L$_{IR}^b$ & 
J$^c$ & H$^c$ & K$^c$ & S$_{5.9}^d$ & S$_{7.7}^d$ \\
\hline
IRAS 12112+0305SW & 0.072 & 0.06 & 0.29 &  5  & 5.8 &1.1& --  & --   & 14.3 & 5.9 & 7.7 \\
IRAS 12112+0305NE & "     & 0.04 & 0.21 & 3.5 & 4.2 &0.8& --  & --   & 14.5 &  "  &  "  \\
IRAS 14348-1447S & 0.082 & 0.07 & 0.33 & 4.5 & 4.7  &1.3&15.6 & 14.7 & 13.7 & 1.3 & 1.9 \\
IRAS 14348-1447N &  "    & 0.03 & 0.17 & 2.4 & 2.4  &0.7&16.1 & 15.1 & 14.3 &  "  &  "  \\
IRAS 17208-0014A & 0.043 & 0.2 & 1.7 & 31.1& 34.9 &2.3&14.0 & 12.9 & 12.4 & 3.2 & 4.9 \\
%IRAS 17208-0014B &  "    &  "  &  "  &  "  &  "   &&     &      &      &     &     \\
IRAS 19254-7245S & 0.062 & 0.19 & 1.15 & 5.3 & 5.6 &1.1&14.0 & 12.7 & 11.7 & 7.0 & 9.5 \\
IRAS 19254-7245N &  "    & 0.01 & 0.05 & 0.2 & 0.2 &0.2&     &      &      &     &     \\
IRAS 20100-4156  & 0.130 & 0.1 & 0.3 & 5.2 & 5.2  &3.6&15.0 & 14.3 & 13.7 & 1.4 & 2.1 \\
IRAS 20551-4250  & 0.043 & 0.3 & 1.9 & 12.8& 10.0 &1.1&13.4 & 12.2 & 12.2 & 4.5 & 7.0 \\
IRAS 22491-1808  & 0.078 & 0.1 & 0.6 & 5.4 & 4.5  &1.5&14.7 & 14.0 & 13.6 & 1.5 & 4.4 \\
IRAS 23128-5919N & 0.044 & 0.05 & 0.3 & 2.9& 2.7 &0.2&14.7 & 13.7 & 13.3 & 4.2 & 6.4 \\
IRAS 23128-5919S &  "    & 0.15 & 1.3 & 7.9& 7.3 &7.0&14.3 & 13.4 & 12.6 &  "  &  "  \\
\hline
\end{tabular}}
\caption{
%\footnotesize{
Photometric NIR to FIR data for our sources. In five cases we
report NIR photometry for two components of the sources. These are the cases where we
were able to disentangle the two components in our L-band observations. 
For these objects we scale the total IRAS fluxes according to the K magnitudes
of the single nuclei.
%All the other data are relative to the total emission of each source. 
Notes: $^a$: flux density in the
IRAS filters, in units of Jy; $^b$: Total infrared luminosity in units of $10^{12}~L_\odot$, 
estimated from Eq.~1 and
assuming the concordance cosmology ($H_0, \Omega_m, \Omega_\lambda)=(70,0.3,0.7)$, 
\citealt{spergel}).
$^c$: Infrared magnitudes from \citet{duc97}, except for the K magnitudes of the two nuclei of 
IRAS~12112+0305, which are from \citet{kimvs}; $^d$: Mid-IR flux densities obtained
with ISO, from \citet{rig99}.}
%}
\end{table*}

%However, the limit is not stringent, for it depends on the assumed cosmology
%(mainly the $H_0$ value, all the sources being at low redshift) and on the 
%nature of the objects under study: objects with a slightly lower luminosity 
%than the above limit, like NGC~6240, still have all the physical properties
%defining a ULIRG, i.e. an infrared emission dominating the bolometric luminosity,
%and largely in excess of the IR emission of normal galaxies (by comparison,
%the Milky Way has an estimated total infrared luminosity in the range 
%$L_{IR}\sim10^{10}~L_\odot$).
%As a consequence, from now on we will refer to all the source in our sample,
%and listed in Tab.~1, as ULIRGs.

Two different considerations drove us in choosing this sample:\\
1. These are the IR brightest ULIRGs, therefore we expect to obtain
the best possible quality for our L-band spectra. This is crucial at the present 
level of development of  L-band diagnostics: as briefly discussed
in the introduction, we still need to perform
detailed studies of bright objects in order to better understand and calibrate
the physical diagnostics, before applying this technique to a larger sample
of fainter sources.\\
2. The sources in our sample are the best studied ULIRGs at other wavelengths.
In particular, complete near-IR photometry and spectroscopy and ISO mid-IR
spectroscopy is available, and in the X-rays we recently performed a complete X-ray
spectroscopic study with {\em XMM-Newton} (\citealt{franc03}, \citealt{braito03}, 
\citealt{braito04}). This will allow us (a) to investigate multi-wavelength indicators
of the AGN/starburst contribution to the IR emission, and (b) to compare
our results in the L-band with existing AGN/starburst classifications based
on observations at other wavelengths.

Out of the 16 ULIRGs in our ULIRGs sample, we selected 
eight sources visible from Chile, and with no previous L-band spectroscopic observations.
For six more sources (IRAS 05189-2524, MKN~231, MKN~273,
UGC~5101, Arp~220, IRAS~08572+3915) 
in the sample 3-4$\mu$m spectra obtained at the 3.6m UKIRT and 3.0m IRTF telescopes
have been published by \citet{im-du} 
 and \citet{im5101L}. 
We will include their results in the discussion in Sect.~4.

For one source (NGC~6240) with an L-band spectrum available in the literature
(\citealt{im-du}), we recently obtained a new, higher quality spectrum with ISAAC.
The results of this observations will be published elsewhere (Risaliti et al.~2005, in prep.).
In the following we will refer to the results of this new observation.

For one source (IRAS~19254-7245S) we already published a detailed spectral analysis
(\citealt{risa03}). The results of this
analysis will be included for completeness in tables and plots, but will not be further discussed.

We report in Tab.~1 the basic known photometric data of our sources.
\section{Observations, and data reduction}

The L-band spectra were obtained 
with the instrument ISAAC 
%(Moorwood et al. 19XX) 
at the UT1 unit of the
Very Large Telescope (VLT) on Cerro Paranal, Chile. 
All but two objects  were observed during three nights on
June 02-04, 2002.
IRAS~20100-4156 and IRAS~22491-1808 were observed on July~30, 2004. 

The sky conditions were photometric, with a seeing of the order of 1 arcsec in K (0.7~arcsec
for IRAS~20100-4156). 
For all but one observations a 20x1~arcsec slit was used. For one source (IRAS~20100-4156)
we used the 20x0.7 slit.
We note that the seeing in the L band is typically better than in K. This is confirmed
by the profile analysis made on our observations (see below for details), which show
that in all cases the fraction of slit loss is lower than 10\%.   
%The K band images obtained from the 2MASS digital archive are shown in Fig.~1,
%with an image of the slit used for the spectroscopic observations.
%We note that in general the double sources are better separated in
%the L-band than they appear in the K band. 
%This is due to both to the higher extended host galaxy emission in K,
%and to a better seeing in our L-band observations, with respect to the average
%seeing in the 2MASS survey.
%
%\begin{figure*}
%\includegraphics[width=17.5cm]{ulirgs_fig1.eps}  
%\label{fcharts}
%\caption{K-band images of our sources from 2MASS. An image of the slit used for
%spectroscopic observations is shown. The area of each figure is 40x40 arcsec.}
%\end{figure*}
%
6 out of 8 sources (the exceptions being IRAS~20551-4250 and IRAS~22491-1808) 
are known mergers, with two distinct
nuclei resolved in infrared imaging. For all these objects we chose the
slit angle in order to obtain spectra for both nuclei. The angles were chosen based
on 2MASS images, and on the short images taken before each spectral observation. 
In three cases (IRAS~12112+0305, 
IRAS~14348-1447, IRAS~23128-5919) we were able to obtain high quality spectra
of both nuclei. In one object (IRAS~19254-7245) we
obtained a high quality spectrum for the brightest nucleus, and a 
low S/N spectrum for the faintest one, enough for an estimate of the flux 
and continuum slope.
In two objects (IRAS~17208-0014 and IRAS~20100-4156) 
we obtained a spectrum of the brightest nucleus only. 
Out of the two single nucleus sources, we obtained a high quality
spectrum of IRAS~20551-4250, while IRAS~22491-1808 was not detected.
%The latter case (the only complete non-detection among the sources in our
%sample) is not surprising, given the faintness of this source in
%the H and K bands (\citealt{duc97}) with respect to the other sources in the
%sample.

The on-source time varies from one hour for the brightest objects to
$\sim3$ hours for the faintest. The observations were performed with the
low resolution grating (LR) in the L band, with a coverage of the $\sim2.9-4.1~\mu$m
wavelength band with a resolution  $\lambda/\Delta\lambda\sim360$.

The main data regarding the observations are listed in Table~2.

\begin{table*}
\centerline{\begin{tabular}{lccccccc}
\hline
Source & Observation Date & T$^a$ & Slit width & Pos. Angle$^b$ & Standard Star & Type & L$^c$ \\
\hline
IRAS 12112+0305   & 2002-06-02 & 180 & 1" & 35 &  Hip 076126 & B3V & 5.92 \\
IRAS 14348-1447   & 2002-06-03 & 180 & 1" & 25 &  Hip 065630 & B3IV& 6.52 \\
IRAS 17208-0014   & 2002-06-02 & 60  & 1" & 0  &  Hip 076126 & B3V & 5.92 \\
IRAS 19254-7245/1 & 2002-06-02 & 60  & 1" & 170 &  Hip 076126 & B3V & 5.92 \\
IRAS 19254-7245/2 & 2002-06-03 & 60  & 1" & 170 &  Hip 000183 & B4V & 5.42 \\
IRAS 20100-4156   & 2004-07-30 & 90  & 0.7"& 0 &  Hip 011407 & B5IV& 4.65 \\
IRAS 20551-4250   & 2002-06-02 & 60  & 1" & 0  &  Hip 076126 & B3V & 5.92 \\
IRAS 22491-1808   & 2002-06-02 & 60  & 1" & 0  &  Hip 076126 & B3V & 5.92 \\
IRAS 23128-5919/1 & 2002-06-02 & 60  & 1" & 0  &  Hip 038593 & B2V & 5.98 \\
IRAS 23128-5919/2 & 2002-06-03 & 60  & 1" & 0  &  Hip 000183 & B4V & 5.42 \\
\hline
\end{tabular}}
\caption{
%\footnotesize{
Observation log of our ISAAC program. 2 sources
were observed twice. Notes: $^a$: observing time in minutes. $^b$:
Slit position angle, with respect to the north-south inclination. The angle increases in counter-clockwise direction, starting from the north. $^c$: L magnitude
of the standard star.}
%}
\end{table*}

In order to avoid saturation due to the high background ($\sim3.9$~mag per arcsec$^2$ in
L-band) all the spectra were taken in chopping mode, with single exposures
of 0.56~s. 
The spectra were then aligned and merged into a single image. We performed
a standard data reduction, consisting of flat-fielding, background subtraction, 
and spectra extraction, using the IRAF~2.11 package.

For each object we also acquired the spectrum of a standard star
immediately before or after the target observation. The star was chosen in order to have
the same airmass as the target, within 0.1, and with an L magnitude between
4 and 6. This allowed us to obtain high quality spectra with 2 minute long
observations. 

Corrections for sky absorption and instrumental response were obtained 
from the standard stars spectra. All the standards are B stars, therefore
we assumed a pure Raleigh-Jeans intrinsic spectrum in the L-band.

In order to facilitate the background subtraction, 
the observations were performed using a mosaic consisting
in a shift of the slit by 20 arcsec up and down along the chosen 
direction. 
However, the instrumental response is not constant along the slit direction
(this effect is mostly removed by the flat-fielding; however, given the
high level of noise, residual differences would strongly affect the final result).
We compensated for this effect performing a separate reduction for
each of the three spectra relative to the three different positions along the slit.
The results before the divisions by the three standard star spectra (obtained 
in the same way) were in general quite different one from the other, both in
absolute flux and in spectral shape. However, after the division by the stars,
the three spectra of all sources overlapped within the errors. 
This gives us an important self-consistency check for our spectral reduction 
procedure. 
%We then
%used the merged spectra for the subsequent analysis. 

In order to obtain a precise absolute calibration, we took into account
aperture effects by analyzing the profiles of both the targets and the
standard stars along the slit. We assumed a Gaussian profile and we estimated
the fraction of flux inside the slit assuming a perfect centering.
We estimate this procedure to have an error of the order of $\sim10-15$\% or
better. In three cases (Table~2) this is confirmed by the comparison 
of the spectra obtained in two different nights, and corrected with two different
standard stars: the final calibrated spectra are always consistent within
the errors. 
A further consistency check is provided by the comparison 
with the magnitudes
at longer wavelengths (when available from ISO observations), which show a good agreement
with our results.
% (see Section XX for details). 

For each source the final spectrum, obtained (a) merging the three single spectra 
relative to the three slit positions, and (b), when two or more observations
were present, merging the final calibrated spectra, was rebinned in order
to have a statistical significance of each bin of at least 2$\sigma$.

The error bars were estimated from the Poissonian noise in the sky counts,
which are by far the dominant source of noise (for ease of comparison,
our brightest source has a magnitude L$\sim$11, i.e. a $\sim600$ times
lower flux than the sky background in the same extraction region).

\begin{figure*}
\includegraphics[width=18cm]{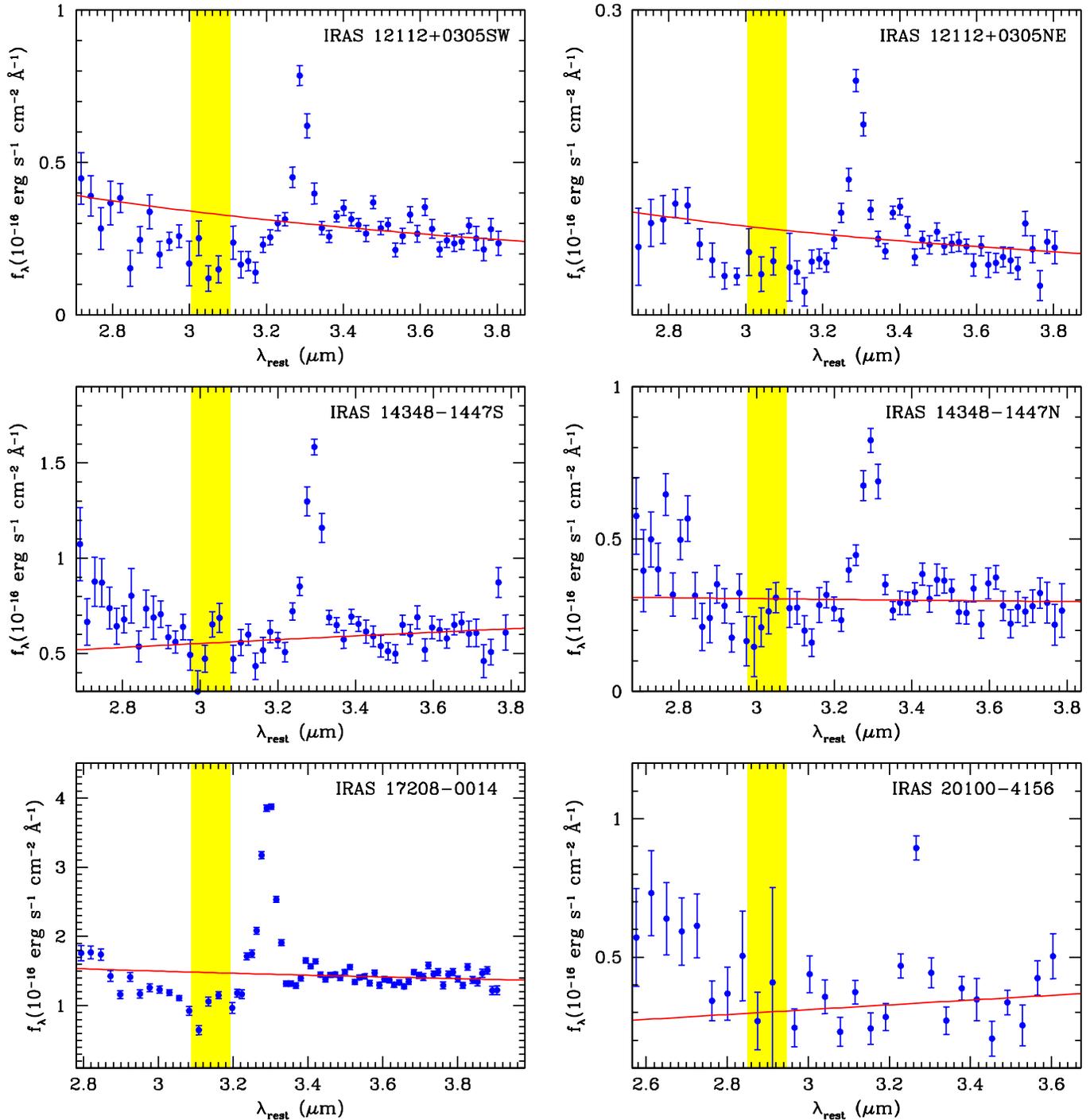}
\label{plot1}
\caption{Flux-calibrated L-band spectra of our sample of ULIRGs. Wavelengths
are in the rest frame. The shaded vertical band shows the spectral region with
bad atmospheric transmission. In each spectrum, the continuous line represents
a power law model fitted
on the spectral region red-ward of the broad emission
feature at 3.3$\mu$m (i.e. at $\sim\lambda>3.4\mu$m in the rest frame).
Exceptions are IRAS~19254-7245S and IRAS~20551-4250, where the two obvious
absorption features at $\sim3.4-3.5\mu$m were also excluded from the fitting region.}
\end{figure*}
\addtocounter{figure}{-1}

\begin{figure*}
\includegraphics[width=18cm]{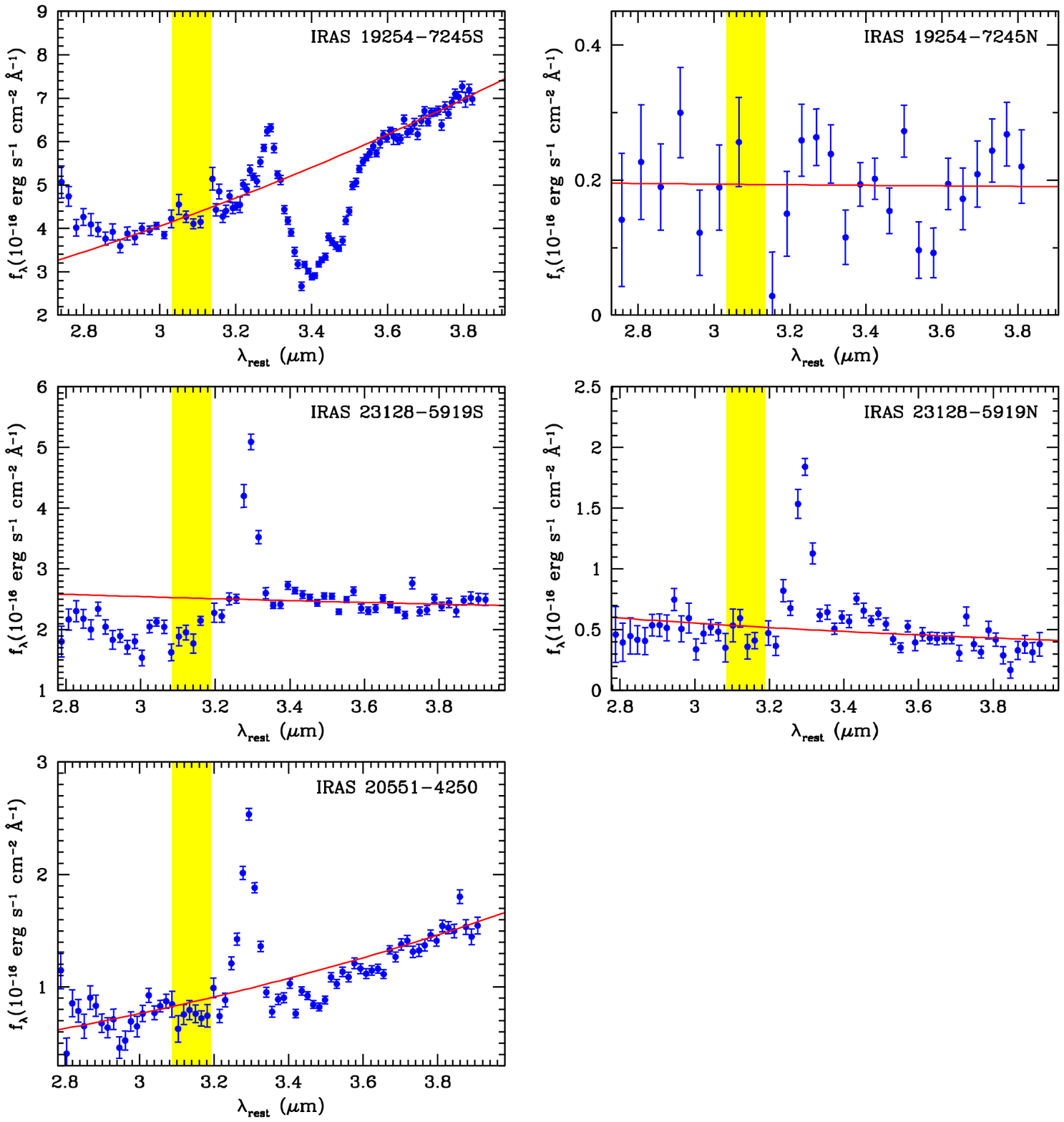}
\caption{- Continued.}
\end{figure*}

The final L-band spectra of our sources are shown in Fig.~1.
%\ref{plot1}.

\section{Data analysis}

As a first step for the interpretation of the L-band spectra of our sample of ULIRGs,
we fitted the ``feature-free'' continuum with a single power law. The chosen wavelength
interval is the 3.4-4~$\mu$m (rest frame) for all sources except IRAS~19254-7245S
and IRAS~20551-4250, where we took into account the conspicuous absorption features 
at $\sim3.4~\mu$m. The best fit power laws are plotted in Fig.~1, in order to better
analyze the absorption and emission features.

As a second step, we fitted the whole spectra in order to quantitatively measure the
relevant parameter of the major spectral features. The best fit model consists of
a power law, an emission line at rest frame wavelength $\lambda_{rest}\sim3.3~\mu$m,
and when needed an absorption feature (with Gaussian profile) 
at $\lambda_{rest}\sim3.4~\mu$m (hydrocarbon dust grain absorption) and one
at $\lambda_{rest}\sim3.1~\mu$m (water ice absorption). 
%Absorption optical depths
%are estimated as the natural logarithmic ratio between the peak values of the best 
%fit Gaussians and the continuum at the peak energy.}\
The results of the spectral fits are shown in Tab.~3.

\begin{table*}
\centerline{\begin{tabular}{lcccccccc}
name              & $\Gamma^a$    & $EW_{3.3}^b$&$\tau_{3.1}^c$&$\tau_{3.4}^d$& L$^e$& 
$\log [(\lambda l_\lambda)_{3.5}/L_\odot]^f$ \\
\hline
IRAS 12112+0305SW & -0.64$\pm0.3$  & 78$\pm21$ & 0.5$\pm0.1$  &  --            & 13.4 & 9.5 \\
IRAS 12112+0305NE & -1.4$\pm0.5$   & 97$\pm34$ & 0.5$\pm0.1$  &  --            & 13.8 & 9.3 \\
IRAS 14348-1447S  &  0.55$\pm0.3$  & 98$\pm23$& --           &  --            & 12.7 & 9.9 \\
IRAS 14348-1447N  & -0.13$\pm0.6$  & 100$\pm37$& --           &  --            & 13.5 & 9.6 \\
IRAS 17208-0014   & -0.32$\pm0.1$  & 83$\pm5$  & 0.4$\pm0.05$ &  --            & 11.7 & 9.7 \\
IRAS 19254-7245S  &  2.30$\pm0.03$ & 26$\pm3$  & --           &  0.8$\pm0.05$  & 10.3 & 10.6 \\
IRAS 19254-7245N  &  0.1$\pm1.2$   & --        & --           &  --            & 13.9 & 9.2 \\
IRAS 20100-4156   &  0.85$\pm0.7$  & 92$\pm41$& --           &  --            & 13.3 & 10.0\\
IRAS 20551-4250   &  2.75$\pm0.13$ & 72$\pm9$& --           &  0.3$\pm0.1$   & 12.0 & 9.6 \\
IRAS 23128-5919S  &  -0.2$\pm0.1$  & 43$\pm5$  & 0.45$\pm0.05$&  --            & 11.1 & 10.0\\
IRAS 23128-5919N  & -1.05$\pm0.2$  & 130$\pm22$  & --           &  --            & 12.7 & 9.4 \\ 
\hline
\end{tabular}}
\caption{Spectral analysis results. 
$^a$: Slope of the continuum power law, defined as $f_\lambda\propto\lambda^\Gamma$.
$^b$: Equivalent width of the 3.3~$\mu$m emission feature, in units of nm.
$^c$: Optical depth of the 3.1~$\mu$m ice absorption feature.
$^d$: Optical depth of the 3.4~$\mu$m hydrocarbon absorption feature.
$^e$: L magnitude.
$^f$: luminosity $\lambda l_\lambda$ at 3.5~$\mu$m, in log of bolometric solar luminosity.
}
\end{table*}

Here we briefly discuss the spectrum of each source, estimating the AGN/starburst
contribution and comparing our results with those obtained through the observations
at other wavelengths. In particular we refer to G98 for ISO
mid-IR spectroscopy, and to \citet{franc03} for {\em XMM-Newton}
X-ray spectroscopy.
\\

{\bf IRAS~12112+0305}. The two nuclei in this source have similar L-band spectra.
In both cases the high equivalent width of the 3.3~$\mu$m line, the absence of 3.4~$\mu$m
absorption and the flat spectrum suggest a starburst origin for their 
infrared luminosity. This is in agreement with the classification at other wavelengths:\\
- in the optical, the [OIII]/H$\beta$ ratio, and the low flux of [NII]~$\lambda6583$\AA,
[SII]~$\lambda6716$\AA~ and [OI]~$\lambda6300$\AA~ with respect to H$\alpha$, suggest
a LINER classification (\citealt{veill99opt}).\\
- in the mid-IR the high equivalent width of the 7.7~$\mu$m PAH
emission feature suggests a pure starburst classification.\\
- in the hard X-rays no indication of AGN activity is found.
\\

{\bf IRAS~14348-1447}. The L-band spectrum of both nuclei is dominated by
a high equivalent width 3.3~$\mu$m emission feature (EW$>$100~nm). The continuum is $\sim$
flat in both cases, with an uncertainty in the brightest nucleus (S) due to a possible
presence of a moderate ($\tau\sim0.1-0.2$) ice absorption. If we allow for ice
absorption, the best fit continuum slope is $\Gamma\sim-0.2$, while without
ice absorption $\Gamma\sim0.5$. The values in Tab.~3 refer to the fit to
a power law+emission line model, without absorption features. This model provides
a good representation of the continuum, with a small excess in the 2.9-3.1~$\mu$m,
which could be due to the tail of the K-band emission coming from a broader region than
the L-band emission.
The starburst classification is confirmed by ISO mid-IR spectroscopy 
and X-ray spectroscopy.
\\

{\bf IRAS~17208-0014}. The L-band spectrum is dominated by a strong
PAH 3.3~$\mu$m emission feature. The continuum slope $\Gamma\sim0.3$ is indicative of
a starburst origin of the L-band emission. An ice absorption feature at 3.1~$\mu$m is clearly
detected. A single Gaussian absorption component is significant at
a $5~\sigma$ statistical level. 
This, together with IRAS~23128-5919S, is the best constrained ice
absorption feature among the sources in our sample, 
thanks to the high S/N spectrum which allows a precise 
continuum determination. The estimated optical depth is $\tau_{3.1}=0.4\pm0.05$.
In the mid-IR and X-rays IRAS~17208-0014 appears dominated by starburst activity.
\\

{\bf IRAS~19254-7245}. The high quality spectrum of the southern nucleus show both
a 3.3~$\mu$m emission feature with an equivalent width (EW$\sim$25~nm),
and two clear AGN features, i.e. a steep continuum ($\Gamma=2.3$) and a strong
absorption at 3.4~$\mu$m ($\tau=0.8$). The substructures of the absorption profile
can be used to constrain the chemical composition of the absorbing dust.
A detailed analysis of this spectrum is presented in \citet{risa03}.
The northern nucleus is extremely faint, and a detailed spectral analysis is not possible.
The continuum is flat, and no spectral feature is present. 
The low S/N prevents us from performing any further analysis. A detailed study of the multiwavelength
spectral energy distribution led to the conclusion that the infrared emission of this nucleus
is due to an inactive galactic component \citep{berta}.
\\

{\bf IRAS~20100-4156}.
Only the brightest nucleus of this source is detected. The spectrum (the lowest
S/N among the primary nuclei in our sources) is dominated by the 3.3~$\mu$m emission
line, suggesting a starburst origin of the L-band emission. The continuum slope is 
quite uncertain, due to the low S/N. An argument similar to that for IRAS~12112-0305NE can 
be applied: the continuum can be fitted (a) by a flat power law, with a
residual excess at 2.9-3~$\mu$m, 
or (b) by a steeper ($\Gamma\sim-0.5$) continuum plus an ice absorption feature with 
$\tau_{3.1}\sim0.5$. In both interpretations, 
no clear indication of the presence of an AGN is found, even if we cannot rule
out a possible minor AGN contribution, due to the low statistics available.
The mid-IR spectrum of this source suggests a starburst classification. In the X-rays,
no clear indication of an AGN is found, however the location of this source in the
X-ray diagnostic diagrams of \citet{franc03} is between the typical 
starbursts and AGNs positions, suggesting a possible contribution of an AGN.
\\

{\bf IRAS~20551-4250}. The high quality spectrum of this source is the most puzzling
in our sample. A strong 3.3~$\mu$m emission feature (EW=70~nm) suggests 
a dominance of the starburst contribution. However, a steep and inverted continuum
slope ($\Gamma\sim2.7$) and the presence of a 3.4~$\mu$m absorption feature ($\tau_{3.4}\sim0.3$)
points to the presence of an AGN dominating the continuum emission.
This is a clear example where the simple classification, based on the 
spectral diagnostics described in the previous section, fails. We will
discuss this source further in Section~6, where an improvement of the above
indicators will be discussed.
The mid-IR spectrum of IRAS~20551-4250 shows clear indications of the presence
of an AGN. Similarly, the X-spectrum is typical of a heavily absorbed
AGN-dominated source. The presence of the AGN was also revealed by the analysis
of the polarized optical spectrum \citep{pernechele}.

\citet{risa03} and \citet{mason} showed that a detailed investigation
of the chemistry of the dust grains responsible of the 3.4~$\mu$m absorption
is possible, based on the analysis of the substructure of the 3.4~$\mu$m absorption
feature. Such analysis on the only other source of our sample showing the
3.4~$\mu$m absorption feature, IRAS~19254-7245, led to the conclusion that
the dust composition in that source is remarkably similar to that observed 
in Galactic stellar sources.

\begin{figure}
\label{i20551}
\includegraphics[width=8cm]{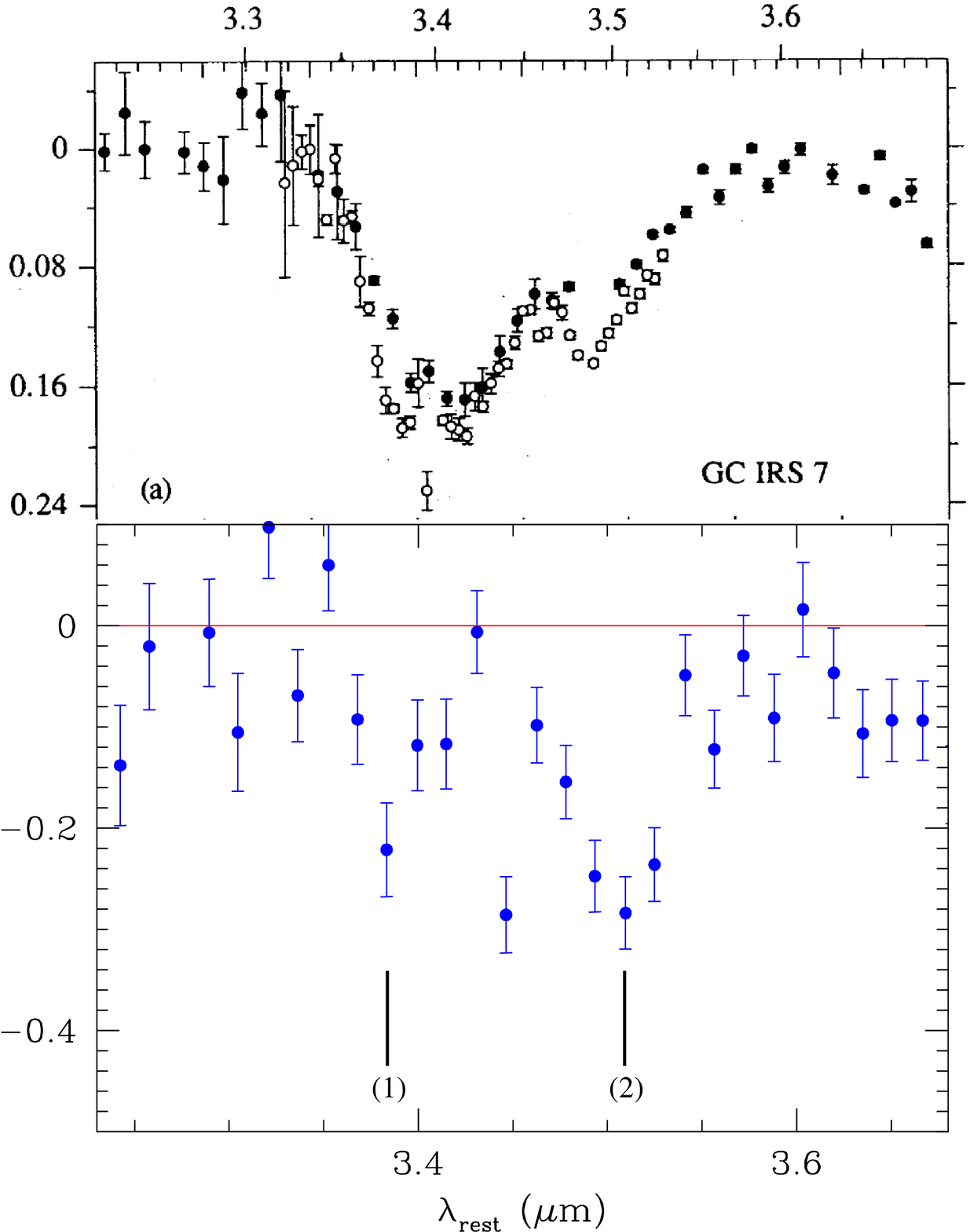}
\caption{Continuum-subtracted spectrum of IRAS~20551-4250 in the 3.3-3.8~$\mu$m
wavelength range, compared with the spectrum of the Galactic center source IRS~7
\citep{pend94}. The complex structure
of the hydrocarbon dust absorption feature is clearly visible. The first dip
is due to vibrational transitions of pure aliphatic hydrocarbon molecules; the second 
dip is due to the same transitions in hydrocarbon molecules where electronegative
group, such as -NO$_2$ or -OH, are present.
}
\end{figure}

The continuum-subtracted spectrum of IRAS~20551-4250 in the 3.3-3.8~$\mu$m range
is plotted in Fig.~2.
A hydrocarbon dust absorption feature is
clearly visible in Fig.~2, and confirmed by the fit with a single Gaussian
absorption component, which is significant at $>3~\sigma$.
A detailed analysis of the chemical
composition of the absorbing dust (analogous to that done in R03 
and \citet{mason} is not possible here, because of the low S/N and
the low optical depth ($\tau\sim0.3$ when fitted with a single Gaussian profile).
However, from a qualitative analysis of Fig.~2 we can tentatively note a 
stronger absorption dip at $\sim3.5~\mu$m with respect to the Galactic source IRS7 
(\citealt{pend94}) and IRAS~19254-7245 (\citealt{risa03}). This would imply 
that the hydrocarbon molecules responsible for absorption in this spectral
region are richer than in the other two sources of electronegative groups such as 
$-NO_2$ and $-OH$.
\\

{\bf IRAS~23128-1808}.
The southern nucleus has a flat continuum, with a moderate equivalent
width 3.3~$\mu$m emission (EW$_{3.3}=43\pm5$, well below the typical 
value for starbursts). A strong ice absorption ($\tau_{3.1}=0.45$) is
clearly detected. The relatively low EW of the 3.3~$\mu$m suggests
the presence of an AGN. The 3.1~$\mu$m absorption feature also suggests
the presence of a compact source. In this case, the good continuum
determination leaves no doubts on the interpretation of the 2.9-3.3~$\mu$m
spectrum, since the alternative scenario of an inverted continuum and
a 2.9-3~$\mu$m excess is ruled out by the data.
The northern nucleus has a typical starburst spectrum with a steep continuum
and a strong 3.3~$\mu$m emission line (EW$\sim130$~nm).
No AGN indication is found in mid-IR spectra, while the hard X-ray spectrum
clearly reveals the presence of an AGN. This source is therefore a
case where the L-band analysis proves to be more effective than
ISO spectroscopy in revealing the presence of an active nucleus as
the origin of the infrared emission.\\

{\bf Summary.} We analyzed the L-band spectra of 7 ULIRGs, 
6 of which consisting of two distinct nuclei. All the single nuclei
were detected except the faintest one in IRAS~20100-4156.
The main results can be summarized as follows:
\begin{itemize}
\item A direct evidence for the presence of an AGN (equivalent width 
of the 3.3~$\mu$m emission feature EW$_{3.3}<50$~nm, and/or steep and inverted
continuum with $\Gamma<-1$, and/or 3.4~$\mu$m absorption feature) has been
detected in three cases (IRAS~19254-7245, IRAS~20551-4250, 
IRAS~23128-5919). 
All these three sources have been  classified as AGNs in our
{\em XMM-Newton} X-ray survey of ULIRGs (\citealt{franc03}).
\item In two of the above cases the AGN indicators are not in agreement:
the continuum of IRAS~23128-5919S is not inverted, and  EW$_{3.3}$ of 
IRAS~20551-4250 is typical of a pure starburst.
\item In two more nuclei (IRAS~12112+0305SW and IRAS~17208-0014) 
the detection of a strong 
ice absorption feature at 3.1~$\mu$m is unambiguous. No other AGN
indication is present in these two sources either in the L-band or
at other wavelengths.
\item A possible  absorption feature in the 3.0-3.2~$\mu$m range
is detected in four more
nuclei (IRAS~12112+0305NE, IRAS~14348-1447S and N,
IRAS~20100-4156). 
In these cases, the interpretation as ice 3.1~$\mu$m
absorption is not unique. An alternative fit can be obtained
with a flatter or inverted continuum, plus a 2.9-3.0~$\mu$m excess due
to the tail of the near-IR emission of the host galaxy.
\end{itemize}
\begin{table*}
\centerline{\begin{tabular}{lcccccc}
\hline
Name              & $\Gamma^a$    & $EW_{3.3}^b$ & $\tau_{3.4}^c$& 
$\log [(\lambda l_\lambda)_{3.5}/L_\odot]^d$ & $\log L_{3.3}/L_\odot^e$ & $L_{IR}^f$\\
\hline
UGC 5101        &   5$\pm0.5$     & 35$\pm10$& 0.65$\pm0.05$ & 10.1 & 8.1 & 0.95\\
Arp 220         &  0.05$\pm0.05$  & 82$\pm10$ & --           &  9.3 & 7.6 & 1.5\\
NGC 6240        &  -0.1$\pm0.1$   & 50$\pm10$ & --           &  9.9 & 8.1 & 0.72\\
MKN 273         &   0.7$\pm0.2$   & 35$\pm5$ & --            & 10.1 & 8.0 & 1.4\\ 
MKN 231         &  -0.5$\pm0.1$   &  2$\pm0.05$& $<0.02$     & 11.5 & 8.2 & 3.4\\
IRAS 08572+3915 &   2.4$\pm0.2$   & $<2$     & 0.9$\pm0.05$  & 10.7 & $<7.3$ & 1.4\\
IRAS 05189-2524 &   0.1$\pm0.1$   & 4$\pm1$  & 0.04$\pm0.01$ & 10.9 & 8.1 & 1.5\\
\hline
NGC 253         &   0.0$\pm0.2$   & 120$\pm10$ & --          & 7.4 & 5.8 & 0.014\\
IC 694          &   0.3$\pm0.2$   & 150$\pm10$ & --          & 9.0 & 7.5 & 0.74\\
MKN 463         &  -0.2$\pm0.2$   & $<0.8$     & 0.12$\pm0.01$ & 11.3  & $<7.6$ & 0.58  \\
IRAS 20460+1925 &  -0.3$\pm0.2$   & $<0.5$     & --          & 11.7 & $<7.9$    & 3.3  \\
IRAS 23060+0505 &   0.1$\pm0.2$   & $<1$       & --          & 11.8 & $<8.3$    & 2.8  \\
\hline
\end{tabular}}
\caption{Spectral parameters of the ULIRGs in the complete sample  
and of five control sources (in the last five lines).
Data are from
\citep{im-du} except for UGC~5101 (\citealt{im5101L}), MKN~463 (\citealt{im-sb}),
and NGC~6240 (Risaliti et al.2005).
$^a$: Spectral index
of a power law fitting the 3.4-4.0~$\mu$m continuum. The values are
inferred from a visual inspection of the published spectra. $^b$: Equivalent width
of the 3.3~$\mu$m emission feature. The errors are conservative estimates
based on visual inspection, except for the source UGC~5101, for which the
error is reported in \citep{im5101L}. $^c$: optical depth of
the 3.4~$\mu$m absorption feature. $^d$: Log of the $\lambda f_\lambda$ luminosity at
3.5~$\mu$m, in units of $L_\odot$. $^e$: Logarithmic ratio between the
luminosity of the 3.3~$\mu$m emission feature and the total IR luminosity. $^f$:
Total infrared luminosity in units of $10^{12}~L_\odot$, estimated from IRAS fluxes
according to Eq.~1.}
\end{table*}

%\begin{table}
%\centerline{\begin{tabular}{lcccc}
%\hline
%Name & Class$^a$ & Ref & $L_{IR}^b$ \\
%%Name              & $\Gamma^a$    & $EW_{3.3}^b$ & $\tau_{3.4}^c$& 
%%$\log [(\lambda l_\lambda)_{3.5}/L_\odot]^d$ & $\log L_{3.3}/L_\odot^e$ \\
%\hline
%NGC 253        &  SB  &  1 &  10.3 &   \\
%IC 694         &  SB  &  1 &  11.7 &   \\
%Mkn 463         & AGN &  2 &  11.8 &   \\ 
%IRAS 24060+1925 & AGN  & 1 &  12.8 &   \\
%IRAS 23060+0505 & AGN  & 1 &  12.8 &   \\
%\hline
%\end{tabular}}
%\caption{Additional bright sources with clear AGN or starburst dominance in
%the L-band. Notes: $^a$: Main energy source: starburst (SB) or AGN;
%$^b$: Infrared luminosity in units of $\log L_\odot$.
%References: 1: \citealt{im-du}; 2: \citealt{im-sb}.}
%\end{table}

\section{Analysis of a complete sample}

The L-band spectra described above 
%provide a powerful tool to investigate the nature of ULIRGs, and 
represent a significant step forward
in our study of ULIRGs, for three main reasons:
\begin{itemize}
\item {\bf High quality}. The high signal-to-noise allowed a precise
determination of the parameters of the main spectral features 
in most spectra (Tab.~3). 
\item {\bf Completeness}. Our small sample consists of the sources of the G98 sample 
visible from Chile at the time of the observations. 
Since the G98 sample is made by the brightest known ULIRGs,
and we did not introduce any physical bias, 
our sources can be considered as representative of ULIRGs in the local Universe.
Moreover, combining our results with those of \citet{im-du}
and \cite{im5101L} we have good quality L-band spectroscopy of
13 objects out of the 15 of the G98 sample. This almost complete sample
can be used to obtain statistical estimates on the average properties
of ULIRGs in the local Universe.
\item {\bf Broad spectral coverage}. The availability of observations
at other wavelengths (optical, near-IR, mid-IR, X-rays) allows a direct
comparison of our results with those obtained with other, independent
spectral diagnostics. This is fundamental in two respects: (a) it
allows a precise calibration of the L-band starburst/AGN indicators,
and (b) a study of the complete spectral energy distribution of these
sources is possible.
\end{itemize}

We will use these properties in order to 
discuss three fundamental issues about the nature of ULIRGs:
(1) Are L-band indicators capable to disentangle the AGN and
starburst contributions in all cases~? And is L-band spectroscopy
effective in finding AGNs in ULIRGs, with respect to indicators
at other wavelengths? (2) How common are AGNs among ULIRGs~?
(3) When an AGN is present, is it possible to estimate the relative
contribution of the AGN and starburst components to the total luminosity~?

In the following, when the sample completeness is required (as,
for example, in estimating the fraction of ULIRGs hosting an AGN)
we will use the complete sample described above.
When completeness is not required, we will add to the above sample
a few more sources, with luminosity in the ULIRG range, or slightly
lower, well known to be either starburst- or AGN-dominated in the L-band.
This will increase the statistics and provide safer references for
our interpretation.
The relevant properties of these ``control sources'' are listed in
Table~4.

\begin{table*}
\centerline{\begin{tabular}{lcccc|cc|cc|cc}
\hline
Name & \multicolumn{4}{c}{L-band} & Optical &Ref & Mid-IR &Ref & X-rays& Ref \\
& $EW_{3.3}$ & $\Gamma$ &$\tau_{3.4}$ &Ref & &&&&&\\
%Optical &Ref & Mid-IR &Ref & X-rays& Ref & \\
\hline
UGC 5101         & AGN  & AGN & AGN&1 & SB &4 & SB &8 & AGN& 9   \\
Arp 220          & SB   & SB  & SB &2 & SB &4 & SB &8 & AGN/SB&10  \\
NGC 6240         & AGN & AGN& SB &13 & SB &5 & SB &8 & AGN&11    \\
MKN 273          & AGN  & AGN & SB &2 & AGN&4 & AGN&8 & AGN&12    \\
MKN 231          & AGN  & SB  & AGN&2 & AGN&6 & AGN&8 & AGN&12    \\
IRAS 05189-2524  & AGN  & SB  & SB &2 & AGN&14& AGN&4 & AGN&15   \\
IRAS 08572+3915  & AGN  & AGN & AGN&2 & SB&4 &  AGN&8 & AGN&6    \\
IRAS 12112+0305  & SB   & SB  & SB &3 & SB &4 & SB &8 & SB &12    \\
IRAS 14348-1447  & SB   & SB  & SB &3 & SB &4 & SB &8 & SB &12    \\
IRAS 17208-0014  & SB   & SB  & SB &3 & SB &5 & SB &8 & SB &12    \\
IRAS 19254-7245  & AGN  & AGN & AGN&3 & AGN&7 & SB &8 & AGN&12    \\
IRAS 20100-4156  & SB   & SB & SB &3 & SB &7 & SB &8 & AGN/SB&12  \\
IRAS 20551-4250  & AGN  & AGN & AGN&3 & SB &7 & SB &8 & AGN&12    \\
IRAS 23128-5919  & AGN  & SB  & SB &3 & SB &4 & SB &8 & AGN&12    \\
\hline
\end{tabular}}
\caption{
Detection of an AGN (either dominant or weak compared 
with the starburst component) in the complete sample of ULIRGs, according to several
L-band, optical, mid-IR and X-ray spectroscopy.
References: 1: \citealt{im5101L}; 2: \citealt{im-du};
3: This work; 4: \citealt{veill99opt}; 5: \citealt{veill95}; 6: \citealt{sand88};
7: \citealt{duc97}; 8: \citealt{lutz-opt}; 9: \citealt{im5101x};
10: \citealt{iwa05}; 11: \citealt{vignati}; 12: \citealt{franc03};
13: \citealt{risa05}; 14: \citealt{laureijs}; 15: \citealt{paolas}}
\end{table*}

%\begin{figure*}
%\label{ind1}
%\includegraphics[width=16cm]{indicators.eps}
%\caption{The four main L-band AGN indicators versus IR luminosity
%for our sample of ULIRGs and a few more ``control'' sources (see text).
%For sources having two resolved nuclei, each nucleus is plotted separately.} 
%\end{figure*}

%The above considerations are reinforced when other L-band
%spectra of non-ULIRG sources are analyzed.
%In Fig.~3 we plot several L-band indicators versus IR luminosity,
%for the ULIRGs in our sample and other well known high luminosity
%(even though not ULIRGs) AGNs
%and starbursts observed in the L-band, and published by Imanishi and
%collaborators. These ``control sources'' are listed in Tab.~6.
%
%The diagrams in Fig.~3 clearly show that no single indicator
%is able to detect the AGN in all sources where we know it is present.

\begin{figure*}
\label{ind2}
\includegraphics[width=17cm]{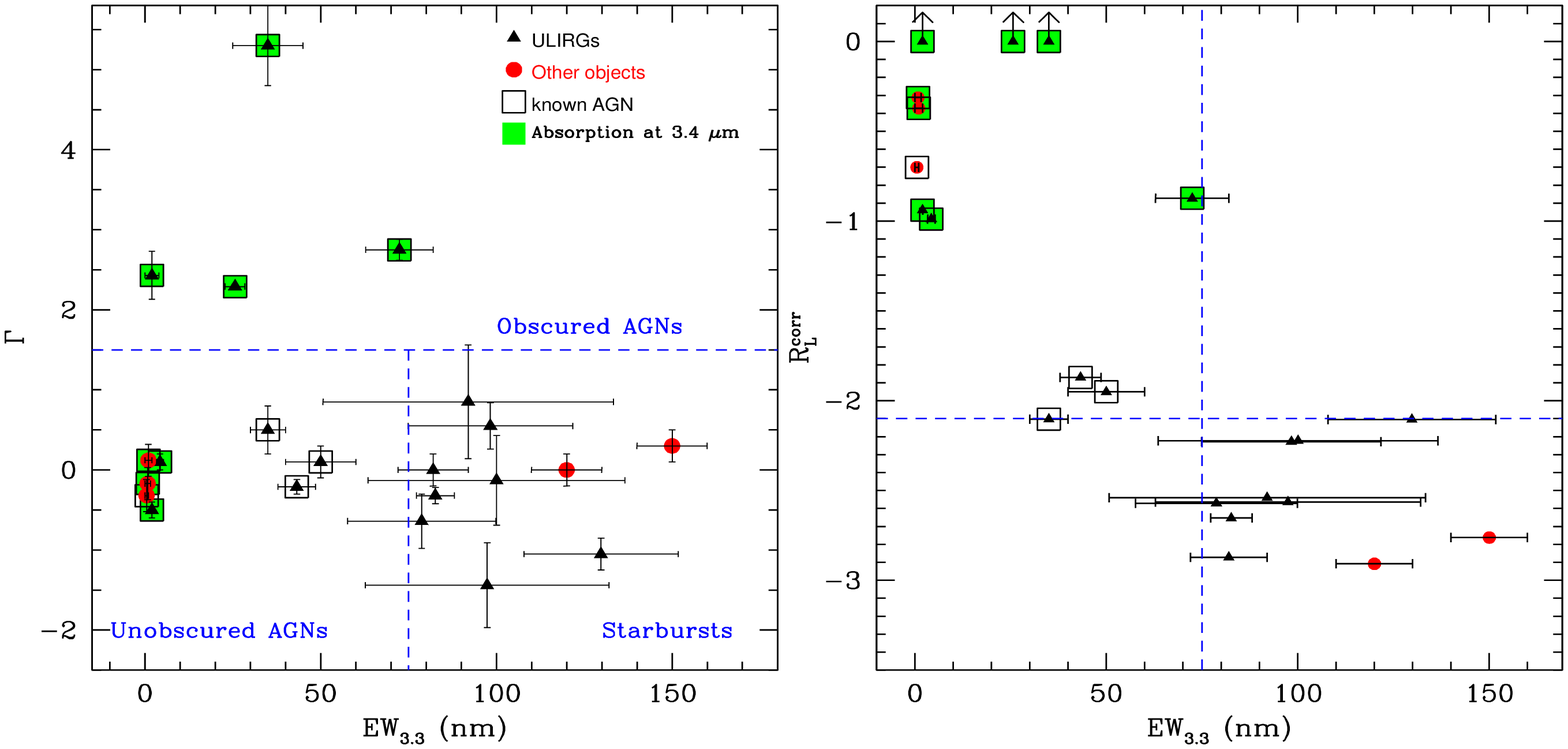}
\caption{Left: 
$\Gamma-EW_{3.3}$ plot for our sample of ULIRGs and control sources.
Right: 
equivalent width
of the 3.3~$\mu$m emission feature versus the logarithmic ratio $R_L^{corr}$
between absorption corrected L-band 
continuum luminosity and total IR luminosity. Errors on $R_L^{corr}$ are
of the order of, or smaller than, the size of the points.
The straight lines are empirical divisions between
AGNs and starbursts. 
} 
\end{figure*}

\subsection{L-band Diagnostic Diagrams}

All the sources in our sample are well studied at other wavelengths,
and the starburst/AGN contribution to their luminosity has been
deeply investigated. The availability of an independent classification
is useful to calibrate our L-band indicators, and to check their 
effectiveness.
In Tab.~5 we report the indications about the presence of an AGN
from several L-band indicators, and from studies at other wavelengths.

The most important result emerging from the analysis of Tab.~3-5 is that
%\begin{itemize}
%\item 
in {\em all cases} where one indicator suggests the presence of
an AGN (i.e. low $EW_{3.3}$, $\tau_{3.4}>0$, or $\Gamma>1$) this is 
confirmed by independent classification.

As a further step in the search for empirical indicators of
the AGN/starburst presence in ULIRGs, we combined the main AGN/SB
indicators in order to study the possible correlations among them.

%two different ways, as shown in Fig.~4\\
a) We plotted our sources in a $EW_{3.3}-\Gamma$ plane (Fig.~3a).
This plot shows that an almost 100\% effective 
L-band classification  is possible: AGNs and starbursts lie in
two different regions in the 
$EW_{3.3}-\Gamma$ plane.\\
b) 
Fig.~3b shows that the same clear division is obtained 
by plotting $EW_{3.3}$ versus the $R_L^{corr}$, defined as the 
logarithmic L-band to total infrared ratio, corrected for L-band continuum absorption.
In order to estimate the absorption in the L-band, 
%in a L/FIR versus $EW_{3.3}$ plane, when the
%L-band/FIR indicator is corrected taking into account the continuum
%absorption inferred from the 3.4~$\mu$m absorption feature.
we made use of the relation $A_L\sim 12\tau_{3.4}$ (\citealt{pend94}).\\
%for the continuum correction.\\

These two results show the power of L-band spectroscopy in detecting AGNs
among ULIRGs. 

It is worth noting that the importance of the above analysis is
related to the possibility of a large extension of the present work.

If we only limit our study to the sample analyzed here,  
our results are only a confirmation of the conclusions already obtained
through studies at other wavelengths, in particular X-rays.
However, the relatively low signal to noise of the X-ray spectra of these objects (which 
are the brightest known ULIRGs) obtained with {\em Chandra} (\citealt{ptak})
and {\em XMM-Newton} (\citealt{franc03}) clearly indicates 
that it is not possible to extend the X-ray analysis to fainter objects.
On the contrary, the quality of our spectra show that it is possible
to obtain a rough estimate of the continuum slope and the 3.3~$\mu$m
PAH feature equivalent width for a source as faint as $L\sim14$. 

Considering the average L-band / FIR ratio (Fig.~3), at least half of 
the sources in the IRAS 1~Jy sample of ULIRGs (\citealt{kim98})
are expected to be brighter than this limit. This implies that our method,
that has been calibrated here thanks to the availability of multi-wavelength
observations of our bright sources, is capable to provide a statistical
analysis on the presence of AGNs in a representative sample of $70$
objects with redshift $z<0.15$. 
(a first sample of $\sim40$ sources observed with Subaru (Imanishi, Dudley \& Maloney~2005)
is already available for this purpose, and more sources are scheduled for observation
with ISAAC at the VLT in the next few months. This will be the subject
of a future paper). 

Even more optimistic predictions are possible if we take into account
that adaptive optics instruments are now available for L-band spectroscopy,
like the NAOS-CONICA at VLT. Since the noise in L-band observations
of our sources is completely dominated by the thermal background ($\sim3.9$~mag/arcsec$^2$)
a factor 10 smaller slit aperture (0.1~arcsec instead of 1 ~arcsec) would
imply a significant increase of the signal-to-noise, even when efficiency
losses due to the adaptive system are taken into account. 

\subsection{Physical properties of the nuclei of ULIRGs}

In the previous Section we discussed the effectiveness of the L-band 
indicators in detecting AGNs in ULIRGs. Even if we demonstrated
that such indicators are indeed effective from a practical point
of view, it is not obvious to find a relation between 
the variety of the observed L-band properties of our sample
of ULIRGs and their physical properties.
%In particular,  we found that: (1) all starburst-dominated sources have
%on a first approximation the same spectral properties, in agreement
%with the expectations based on the physical arguments outlines in 
%Sect.~3; (2) sources hosting an AGN show a remarkable etherogeneity
%in spectral properties. 

Here we want to investigate this fundamental issue, through a
physical interpretation of the indicators discussed above and
of the results of observations at other wavelengths.

\begin{figure}
\label{ind3}
\includegraphics[width=8cm]{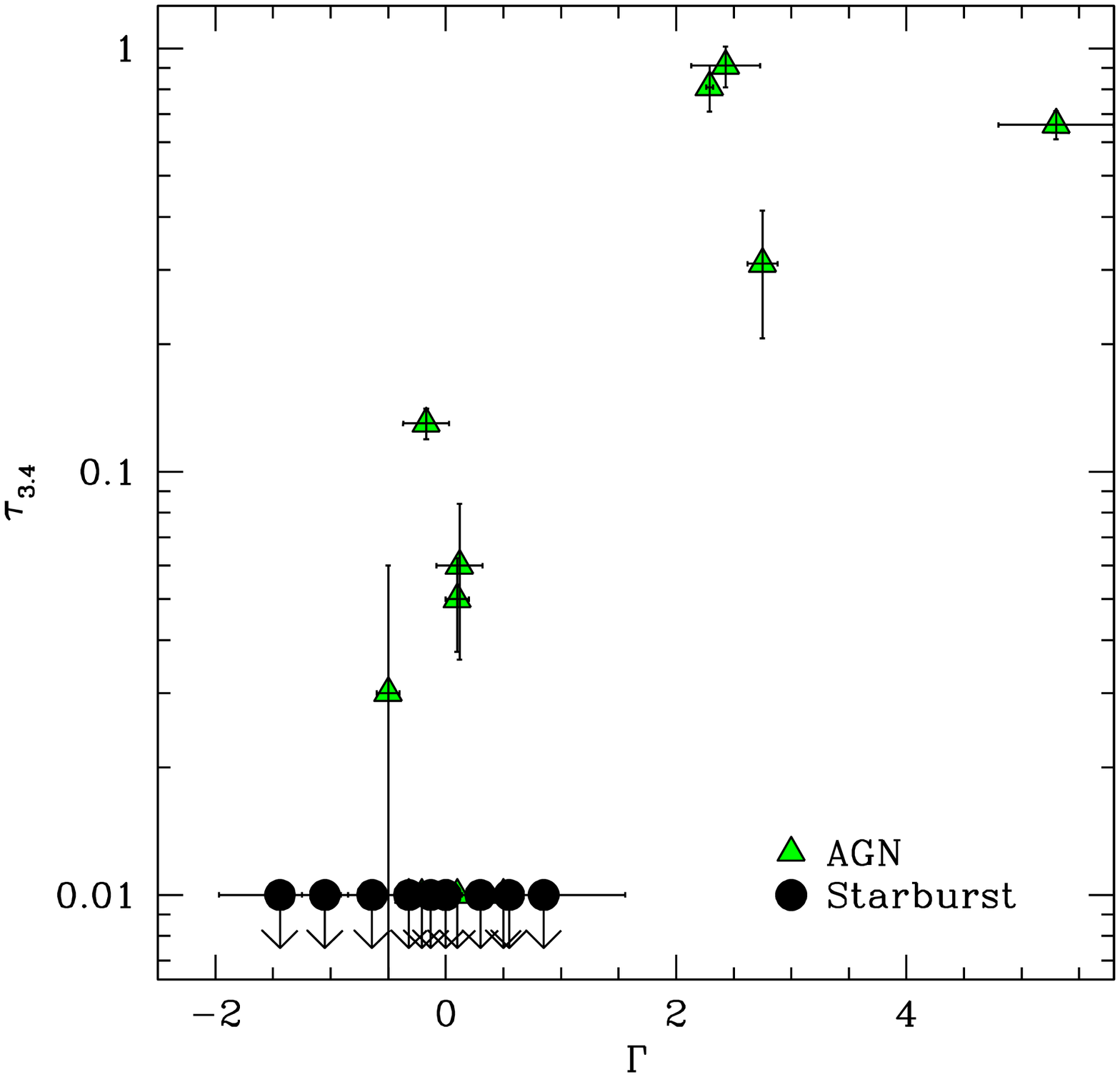}
\caption{3.4~$\mu$m absorption versus continuum slope for
our sample of ULIRGs and control sources. Typical errors on $\tau_{3.4}$
are of the order of 10\%.}
\end{figure}

Among AGNs, an apparent spread in spectral properties suggests
a high heterogeneity in the physical properties of the central emitting
regions. However, two further indications suggest a simple physical interpretation:\\
1) The
correlation between the continuum slope, $\Gamma$, and the optical
depth at 3.4~$\mu$m, $\tau_{3.4}$ is shown in Fig.~4.
The sources with $\Gamma>1$ are those with a high 3.4~$\mu$m absorption
($\tau_{3.4}>0.3$). All the other sources are clustered around
the ($\Gamma\sim0, EW_{3.3}\sim0$) point in Fig.~3a.\\
2) All AGNs in the ($\Gamma\sim0, EW_{3.3}\sim0$) group (Fig.~3)
are not heavily absorbed in the
X-rays (column densities in the range $22 < \log N_H < 23.5$), while
all the high $\Gamma$, high $\tau_{3.4}$ objects have $\log N_H > 24$.
Assuming a Galactic dust to gas ratio, $\log N_H\sim23$ implies $A_L\sim3$.
If the dust to gas ratio is lower than Galactic, as suggested in all
cases where a direct measurement is possible (\citealt{maio}),
the absorption in the L-band is even smaller. \\

Combining the above indications, a physical interpretation clearly emerges:\\
- ($\Gamma\sim0, EW_{3.3}\sim0$) AGNs (Fig.~3) are not heavily obscured in the L-band.
As a consequence, no strong absorption feature is present, and the continuum
emission is due to the direct emission of the hot circumnuclear dust
heated by the optical/UV primary AGN emission. 
In these cases, the AGN component dominates in the L band even when
a significant starburst contribution to the bolometric luminosity is
present. This is due to the $\sim$two order of magnitudes higher L-band/bolometric
ratio in AGN than in starbursts. As a consequence, the PAH emission
feature at 3.3~$\mu$m is either weak or absent.
\\
- ($\Gamma>1, \tau_{3.4}>0.3$) AGNs are heavily obscured in the L-band. The 
observed continuum is the result of a strong dust extinction by dust.
The heavy dust obscuration results into a highly reddened continuum (high $\Gamma$)
and a strong absorption feature at 3.4~$\mu$m.
Since the AGN continuum is heavily absorbed, if a starburst
is contributing to the bolometric luminosity, it is possible that
it also represents a significant fraction of the L-band emission, depending
on the amount of AGN extinction. This explains the wide range of measured values of
the equivalent width of the 3.3~$\mu$m emission feature in these objects.\\
A more quantitative treatment of the relative starburst/AGN emission in ULIRGs
will be discussed in the next Section~6.3.

With regard to starburst dominated sources, the observed
spread in the L-band spectral parameters (continuum slope $\Gamma$,
$EW_{3.3}$) is much smaller than in 
sources known to host an AGN. This spread could be either intrinsic
(due to the age of the stellar populations)
or due to an AGN component which has not been revealed through our
indicators. For example, an unobscured or moderately
obscured AGN contributing $\sim$1/3 of 
the L-band emission would probably be missed by our analysis,
and would imply a change by $\sim30$\% of the measurements of
the above spectral parameters. We note that in order to have
such a case the contribution of the AGN to the bolometric luminosity
has to be negligible (of the order of a percent or smaller).

Finally, we cannot exclude that a powerful AGN, even dominating
the bolometric luminosity, with a high enough obscuration
to make negligible its contribution in the L-band, is present
in some of the starburst-classified ULIRGs. In this case the only
affected parameter would be the L-band to bolometric ratio.
Given the large spread in this parameter, (Fig.~4) even among starburst-classified
sources alone, this possibility cannot be excluded for any of 
the sources in our sample.

\subsection{The AGN contribution to ULIRGs}

Estimating the relative contribution of
AGNs to the luminosity of ULIRGs is a more ambitious goal than
the simple detection of AGNs inside ULIRGs. A multi-wavelength analysis,
based on models of the whole optical to FIR Spectral Energy Distribution
(SED) is probably the most powerful approach to this problem. 
Such a study for several of our sources is currently being
finalized (Fritz et al.~2005, in prep.).

Here we only focus on L-band diagnostics.

Our analysis of the AGN contribution to ULIRGs luminosity
is done in two steps: (a) estimate of the fraction of ULIRGs hosting
an AGN; (b) estimate of the relative AGN/starburst contribution 
to the  total luminosity in sources hosting an AGN.

\subsubsection{ Fraction of ULIRGs hosting AGNs}
Out of the 14 ULIRGs in Table~5, only 5 do not
show any evidence for the presence of an AGN. 
These sources belong to the flux-limited sample
defined by the criterion $S_{60}>5.24$~Jy (Sect.~2).
In order to complete this sample only two objects are missing:
IRAS~15250+3609 and IRAS~22491-1808. The first one has
never been observed in the L-band, while the second one
was not detected in our ISAAC observation. Both sources
have been observed in the X-rays with {\em XMM-Newton}
(\citealt{franc03}), and in the mid-IR with ISO (G98),
and no indication of the presence of an active nucleus has
been found in both cases. Therefore, in order to be conservative in
the estimate of the fraction of AGN-hosting ULIRGs, we classify
these two sources as starburst-dominated.
The final sample of 16 ULIRGs consists of 7 starburst dominated
objects (44\%) and 9 objects with a significant AGN contribution
(56\%). 

These results have been obtained by using a flux selection 
at 60~$\mu$m. This is not an unbiased selection criterion,
because of the
higher fraction of bolometric luminosity emitted at 60~$\mu$m
by starbursts with respect to AGNs (see also the discussion in
\citealt{risa00}). In our sample the average
fraction of the total IR luminosity  emitted at 60~$\mu$m 
(using Eq.~1) is 0.62 for the seven starbursts, and 0.52 for
the 9 AGNs. This implies that in order to have the same
total IR flux limit for AGNs and starbursts, the flux limit
at 60~$\mu$m must be lower for AGNs by a factor (0.62/0.52=1.19).
Assuming a euclidean logN-logS, the expected
correction for the number of AGNs in a 60~$\mu$m flux
limited sample is 1.19$^{3/2}$=1.30.
When this correction is taken into account, the final estimate
of the fraction of ULIRGs hosting an AGN is 63\%.

\subsubsection{Deconvolution of the AGN and starburst components}

An accurate estimate of the relative starburst/AGN contribution
for each source in the L-band requires a spectral decomposition of 
the two components. This is extremely difficult due to the
spread of intrinsic properties shown in Fig.~3 and~4.

A zeroth-order indicator of the relative contribution of the
two components is the logarithmic ratio $R_L^{corr}$ between the absorption-corrected 
continuum luminosity
in the L-band and the total IR luminosity (Fig.~3b).
It is shown in Fig.~3b that 
sources whose L-band spectrum is dominated by the unobscured emission of an AGN
have $R_L^{corr}>-1$, while the average value for
starburst-dominated sources is $R_L^{corr}\sim -2.6$.
For comparison, $R_L=-0.57$ in the quasar spectral energy distribution 
of \citep{elvis94}.
The huge difference between the two values of $R_L^{corr}$ for starbursts
and AGNs, and the relatively large dispersion around the average values makes
an accurate estimate of the relative contributions of the two components
extremely difficult. For example, if 50\% of the total IR luminosity
of a ULIRG hosting an unobscured AGN is due to a circumnuclear starburst,
$R_L^{corr}$ would be changed only by a factor of two, well within the spread
shown in Fig.~3b. The conclusion is that the L-band spectrum of 
ULIRGs hosting an unobscured AGN is dominated by the AGN emission
even if the starburst is the dominant energy source. 

%As a consequence,
%we can only estimate a lower limit to the AGN contribution of 20-30\% 
%based on the comparison between $R_L^{corr}$ in these objects 
%($R_L^{corr}\sim-1$) and in
%local quasars ($R_L=-0.57$). 
%
%The estimate is even more complicated for obscured AGNs. 3 out of 4
%sources hosting a heavily obscured AGN have $\log R_L^{corr}>0$,
%which is physically not acceptable. This suggests a dust composition 
%different from that assumed in the correction factors from observed to intrinsic
%L-band luminosity (see next Subsection). Since the real correction factor
%is not known with precision, and, again, a minor contribution from
%an AGN in terms of bolometric luminosity is enough to dominate the
%L-band emission, no estimates of the relative starburst/AGN contributions
%can be done.

The situation is significantly different for obscured AGNs. In these
cases dust extinction of a few magnitudes can make the AGN emission 
comparable to that of a starburst of similar bolometric luminosity.
The continuum extinction can be estimated from the measured 3.4-4~$\mu$m slope.

In order to quantitatively estimate these contributions, we made several
assumptions on the intrinsic AGN and starburst emission, and developed a
simple model to reproduce the observed L-band spectra.
The main features of the model are the following:
\begin{itemize}
\item The observed L-band spectra are the composition of AGN and starburst
contributions, which are parametrized as:
\begin{equation}
f_\lambda=\alpha f_{\lambda,AGN} e^{-\tau_L(\lambda)} + (1-\alpha) f_{\lambda,SB}
\end{equation}
where $\alpha$ is the fraction of the L-band luminosity due to the AGN,
%$K$ is the ratio between the L-band emission in AGNs and in starbursts with the
%same bolometric luminosity, 
and $\tau_L(\lambda)$ is the wavelength dependent
continuum optical depth in the L-band. $f_{\lambda,AGN}$ and $f_{\lambda,SB}$
are the intrinsic AGN and starburst L-band spectra.
\item Based on the results of the spectral analysis shown in Table~3 and~4, and
in Figure~3 and~4, we assume $f_{\lambda,AGN} = \lambda^{-0.5}$
and  $f_{\lambda,SB} = \lambda^{-0.5} + f_{PAH}$. $f_{PAH}$ is the contribution
of the 3.3~$\mu$m emission feature, normalized in order to have $EW_{3.3,SB}=110~nm$
with respect to the pure starburst continuum. The choice of the spectral
index ($\Gamma=-0.5$ for both pure AGN and pure starburst) 
has been driven by the average values observed in our sample, both for
starburst-dominated and AGN-dominated ULIRGs. The value for AGNs is in agreement with
L-band spectra of pure type 1 AGNs (\citealt{im-wada}). The value for starbursts
is the average from our data, and has a large scatter (Fig.~3a). However, this
uncertainty is not a major problem for our model (see details below).
We note that our pure starburst spectra differ on average from the average spectra
of lower luminosity normal galaxies (which have $\Gamma\sim-2$, \citealt{lu}), probably due
to a larger contribution by hot dust in powerful starbursts.
  
\item We assume an extinction law $\tau_L(\lambda)\propto\lambda^{-1.75}$ 
\citep{cardelli}.
\end{itemize}
The model has two free parameters: $\tau_L$ and $\alpha$,
where $\tau_L=\tau_L(\lambda=3.5~\mu$m). 
For each pair of these parameters we compute the slope, $\Gamma$, and
the PAH feature equivalent width, $EW_{3.3}$, for the composite spectrum.
A good analytic approximation (within a few percent) of these relations can be obtained
noting that the wavelength interval ($\lambda_1,\lambda_2$) used for the continuum fit is
rather narrow ($\lambda_1\sim3.4~\mu$m, $\lambda_2\sim4~\mu$m). Using a first
order expansion in $\delta=(\lambda_2-\lambda_1)/\lambda_1$ we easily obtain:
\begin{equation}
EW_{3.3}=\frac{(1-\alpha) EW_{3.3,SB}}{\alpha e^{-\tau_L}+(1-\alpha)} 
\end{equation}
\begin{equation}
\Gamma=-0.5+\frac{1.75 \tau \alpha e^{-\tau_L}+(1-\alpha)}{\alpha e^{-\tau_L}+(1-\alpha)}
\end{equation}

We plot the results in Fig.~5. The two groups of curves are obtained
freezing one of the two parameters and varying the other. Therefore, they form a
grid which can be directly used to estimate the relative AGN contribution
and the AGN L-band extinction of each single ULIRG.

\begin{figure*}
\includegraphics[width=17.5cm]{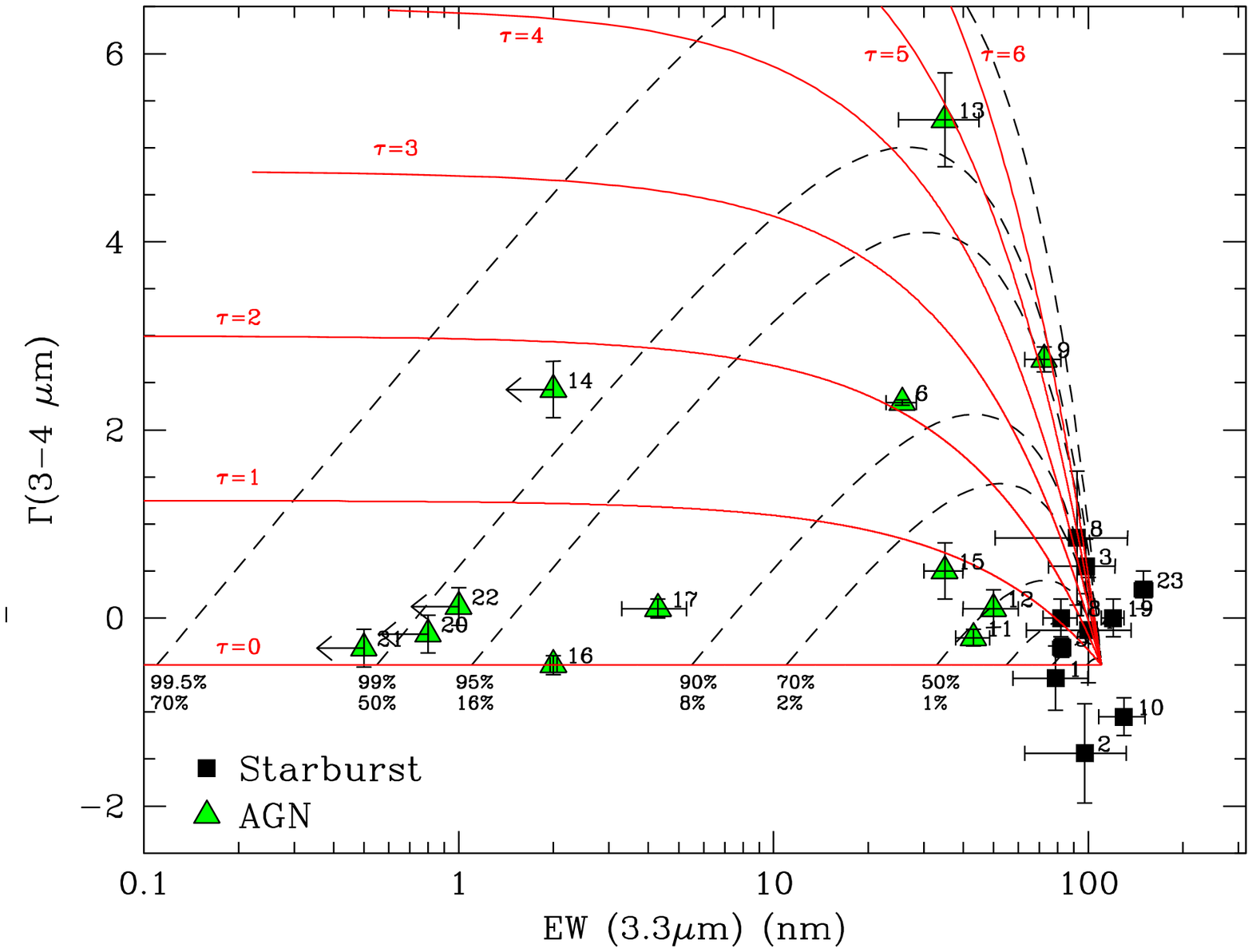}
\label{model}
\caption{Estimates of the relative AGN/SB contribution 
in a L-band continuum slope 
versus $EW_{3.3}$ plot, according to the model described in the text.
Dashed black lines represent combinations of AGN and SB with a fixed
relative contribution with a varying L-band absorption 
of the AGN component (from $\tau_L=0$ to $\tau_L=6$). 
The two numbers below each line are the
fractions (in percent) of the AGN contribution to the L-band
(first number) and to the bolometric emission (second number). 
Continuous red lines represent combinations with a fixed AGN absorption
(as labeled for each line) and a varying relative contribution.
Following a dashed line from the bottom-left, it starts from a no-absorption point
and moves to the upper-right direction as absorption (i.e. reddening AND 
attenuation of the AGN component) increases. It then moves to the lower
right direction because as the reddening further increases,
the starburst contribution becomes more important
than the absorbed AGN component.
Following a continuous line, it starts from a right point representing 
a pure AGN, and then moves to the bottom-left as the starburst contribution increases.} 
Numbers identifying each source refer to Table~6.
\end{figure*}

Equations~3 and~4 can be easily solved in terms of $\alpha$ and $\tau_L$.
A quantitative analysis of the results is shown in the Appendix.
Here we concentrate on a visual analysis of Fig.~5, from which
both the strengths and the limitations of the model
clearly emerge:\\
- It is possible to estimate with good accuracy the AGN contribution to the
L-band emission in a 
large region of the parameter space ($\alpha>0.1, \tau_L<4$).\\
- Our diagnostics fails around the pure starburst region, where all
the model lines converge and high degeneracy among the parameters is present.
Therefore, we cannot exclude that L-band starburst-dominated objects
host a heavily buried ($\tau_L>5$) AGN. However, in the cases where
the AGN dominates, it is still possible to disentangle its contribution
even in highly obscured objects, such as IRAS~20551-4250 and
UGC~5101 (numbers 7 and 8 in Fig.~5). \\
- The dependence on the choice of the intrinsic AGN and SB spectra is not strong:
if, for example, the ``pure starburst'' point is moved (provided that it
remains inside the starburst region of the plot), the only
affected region is the one close to the point itself, where
the model lines converge and the estimate of the parameters is
not reliable. 
These limitations are quantitatively discussed in the Appendix,
through the analysis of the statistical errors in the model.

\begin{table*}
\centerline{\begin{tabular}{llcccc|llcccc}
N & Source & $\alpha$(\%) & $\alpha_{BOL}$(\%) & $\tau_L$ & $\tau_L(\tau_{3.4})$ & 
N & Source & $\alpha$(\%) & $\alpha_{BOL}$(\%) & $\tau_L$ & $\tau_L(\tau_{3.4})$\\
\hline
1 & IRAS 12112+0305SW & 0           & 0         & --  & -- & 
13 & UGC 5101        &  99$\pm1$    & 68$\pm29$ & 4.6$\pm0.9$ & 7.8$\pm2.6$ \\
2 & IRAS 12112+0305NE & 0           & 0         & --  & -- & 
14 & IRAS 08572+3915 &  99.5$\pm0.5$& 75$\pm23$ & 1,7$\pm0.2$  & 10.8$\pm3.6$ \\
3 & IRAS 14348-1447S & 0            & 0         & --  & -- & 
15 & MKN 273         &  79$\pm8$    & 4$\pm1$   & 0.6$\pm0.2$  & -- \\
4 & IRAS 14348-1447N & 0            & 0         & --  & -- & 
16 & MKN 231         &  98$\pm0.5$  & 35$\pm2$  & $<0.05$    & -- \\
5 & IRAS 17208-0014  & 0            & 0         & --  & -- & 
17 & IRAS 05189-2524 &  97$\pm1$    & 26$\pm6$  & 0.3$\pm0.06$  & 0.5$\pm0.2$ \\
6 & IRAS 19254-7245S & 96$\pm1$     & 18$\pm3$  & 1.9$\pm0.1$   & 9.6$\pm3.2$ & 
18 & Arp 220         &  0           & 0         &--   & -- \\
7 & IRAS 19254-7245N & 0            & 0         & --  & -- & 
19 & NGC 253         &  0           & 0         & --   & -- \\
8 & IRAS 20100-4156  & 0            & 0         & --  & -- & 
20 & MKN 463         &  99$\pm1$    & 62$\pm26$ & 0.2$\pm0.1$ & 1.4$\pm0.4$ \\
9 & IRAS 20551-4250  & 97$\pm3$     & 28$\pm30$ & 4.3$\pm1.1$   & 3.6$\pm1.6$ & 
21 & IRAS 20460+1925 &  99.5$\pm0.5$& 71$\pm23$ & 0.1$\pm0.1$    & -- \\
10& IRAS 23128-5919S & 58$\pm8$     & 1.5$\pm0.5$& $<0.2$ & -- & 
22 & IRAS 23060+0505 &  99$\pm1$    & 61$\pm27$ & 0.3$\pm0.1$    & -- \\
11& IRAS 23128-5919N & 0            & 0         &  --  & -- & 
23 & IC 694          &  0           & 0         & --   & -- \\
12& NGC 6240        &  58$\pm15$    & 1.5$\pm0.5$& 0.2$_{-0.2}^{+0.3}$  & -- \\
\hline
\end{tabular}}
\caption{Relative AGN contribution to the L-band luminosity ($\alpha$) and
to the total bolometric
luminosity, $\alpha_{BOL}$, and L-band optical depth for the AGN component, $\tau_L$, 
for the complete sample (first 18 sources) and for the control sources
(last 5 sources), as estimated with the model described in Section~6.4.2, using
the equations in Appendix. The second estimate of the
continuum L-band optical depth, $\tau_L(\tau_{3.4})$, is based on the
optical depth $\tau_{3.4}$ of the 3.4~$\mu$m absorption feature, according
to a standard extinction curve (Pendleton et al.~1994). Quoted errors are
due to measurement uncertainties only. Systematic effects maybe relevant in
$\alpha_{BOL}$ (of the same order as statistical ones for AGN-dominated sources, and
2-3 times larger for starburst-dominated sources). }
\end{table*}

\subsubsection{AGN/SB contributions to the bolometric luminosity}

A further step can be made estimating the AGN contribution to the
bolometric luminosity. This is possible taking into account the
different L-band to bolometric corrections for AGN and starbursts, $R_L(AGN)$ 
and $R_L(SB)$. We introduce a new parameter, $K$, defined as the ratio 
between the two bolometric corrections
From the considerations made above ($R_L(SB)\sim-2.6, R_L(AGN)\sim-0.6$)
we have $K\sim100$.

The fraction of the bolometric luminosity due to the AGN, $\alpha_{BOL}$,
is then given by
\begin{equation}
\alpha_{BOL}=\frac{\alpha}{\alpha+K(1-\alpha)}
\end{equation}

We list the results of the ($\alpha, \alpha_{BOL}, \tau_L$) estimates in Table~6.
%We discuss the continuum absorption in the next Subsection.
%Here we concentrate on
Restricting our analysis on the complete sample (i.e. neglecting the
five ``control sources'') and adding up the contributions of both nuclei
in the double sources, we obtain that 
the relative AGN contribution to the bolometric
luminosity, $\alpha_{BOL}$, is higher than 50\% for 2 out of 9 AGNs.
Correcting for incompleteness as explained in 
Section 6.4.1, we end up with the following conclusions: \\
$\bullet$ Fraction of sources with only a starburst component detected: $\sim40$\%;\\
$\bullet$ Fraction of sources with a dominant starburst component and
a minor AGN component $\sim30$\%;\\
$\bullet$ Fraction of sources dominated by an AGN: $\sim30$\%.\\

Simply averaging the AGN and starburst components we find that the 
overall contribution of starbursts to the luminosity of ULIRGs
is $\sim75$\%, while the AGN contribution is the remaining 25\%.
Considering that the typical errors in estimating $\alpha$ in Fig.~5
vary from a few percent to $\sim50$\%, and the propagation of errors
in the computation of the average, we estimate the statistical uncertainty on
the total averages to be of the order of 5-10\%.

However, the determination of $\alpha_{BOL}$ has several caveats which must be taken 
into account. While the main uncertainties in the determination of $\alpha$ and $\tau$
are due to measurement errors, other systematic effects may be relevant, or even dominant.
In particular, two possible complications must be considered:
\begin{enumerate}
\item A completely hidden component, either AGN or starburst, due to heavy obscuration
in the L-band, cannot be excluded. This would imply a contribution to the bolometric luminosity 
from the FIR, with no indications in the L-band. Such a scenario has been proposed
for instance by \citet{fischer} and \citet{im5101L} to explain the dependence of the PAH-to-FIR
flux ratio with luminosity in starbursts: lower ratios are observed in higher luminosity
sources, suggesting the presence of an obscured component. In order to test this scenario,
we compared the total luminosity predicted with our model (and therefore based on the observed 
L-band components) with the FIR luminosities measured by IRAS.
The results, plotted in Fig.~6 show that in most cases the two values match,
suggesting that hidden components do not represent an important contribution to the bolometric luminosity.
However, several points are significantly off the 1:1 relation. In these cases a contribution from 
a completely obscured component cannot be ruled out. Moreover, while the bolometric correction
used for AGN is independently determined (Elvis et al.~1994), the L-band to FIR ratio for starbursts is
inferred from our sample (Fig.~3b). Therefore, if a completely obscured component is responsible 
of a fraction of the FIR emission, we are implicitely assuming that it consists of a pure starburst.
As a consequence, our estimates of the AGN contribution could actually be lower limits in some cases.

\item The spread in the L-band to bolometric ratio, both in starbursts and in AGNs (Fig.~3b)
makes the determination of the parameter $K$ rather uncertain. Moreover, the assumed average
ratio for AGNs is that obtained from a sample of quasars (\citealt{elvis94}) which could be significantly 
different from that of AGN-dominated ULIRGs, due to possible different amounts of circumnuclear dust.
Taking into account this uncertainty, we conservatively estimate a range for $K$ from $\sim30$ to $\sim300$.
By propagating the error on Equation~5 we obtain that this is the dominant uncertainty on $\alpha_{BOL}$,
at least for the sources not dominated by the AGN contribution in the L-band. For example, in the cases of
NGC~6240 and IRAS~23128-5919S (Tab.~6) we obtain $\alpha_{BOL}(K=30)$=3\%, $\alpha_{BOL}(K=300)$=0.3\%.
Note however that even assuming such large uncertainties, the basic result that in these sources the AGN
total contribution is negligible still holds. We note that our result is not in agreement with that
of \citet{lutz6240} who estimate an AGN contribution in the range 25-50\%. This complex source will be discussed in more detail in a forthcoming paper (\citealt{risa05}).  
\end{enumerate}
Summarizing the above analysis, we conclude that our simple model provides a reliable estimate of the relative AGN/starburst contribution for statistical purposes, with possible exceptions in single sources due to heavily obscured AGN or starburst components.
\begin{figure}
\includegraphics[width=9cm]{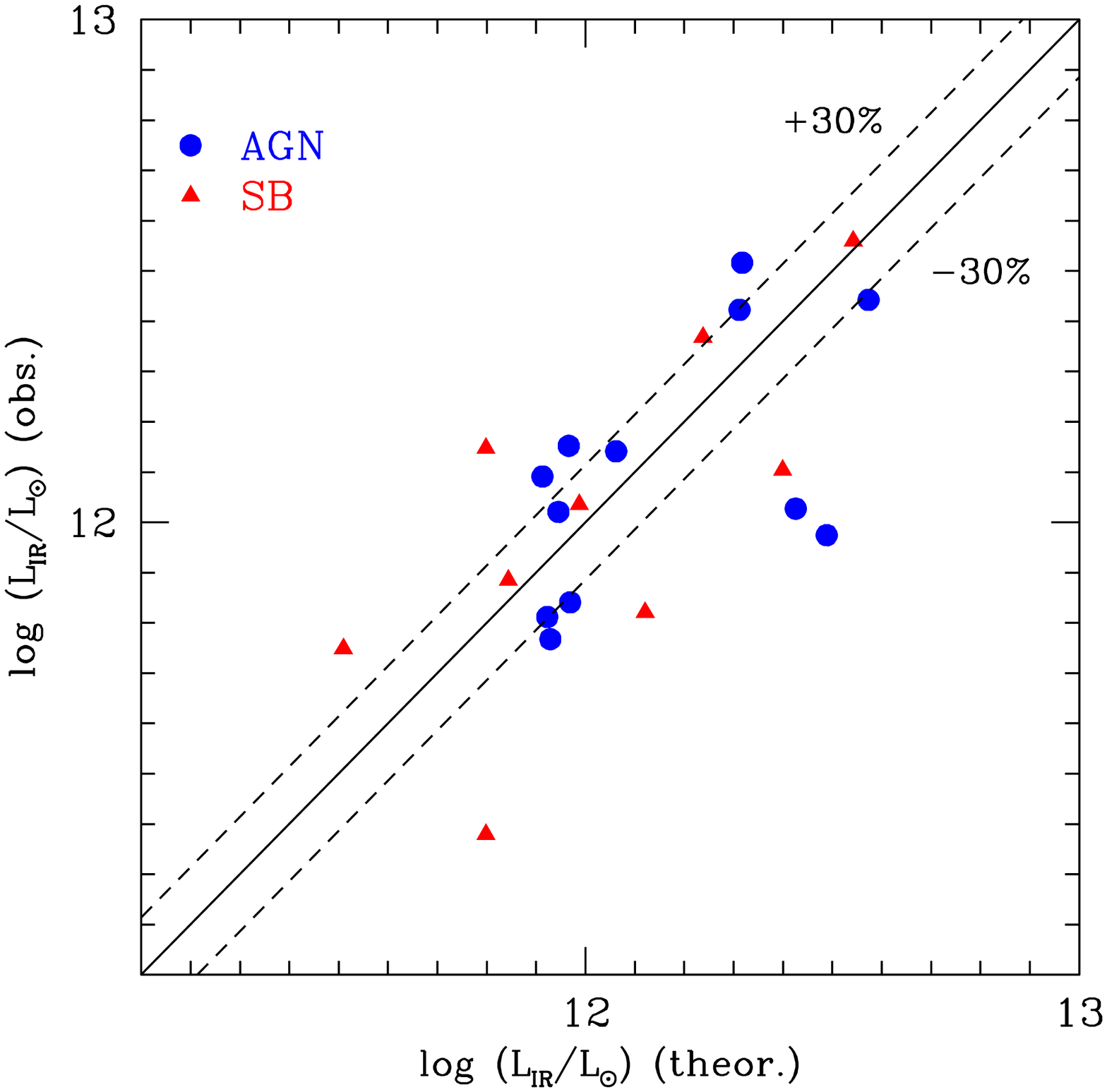}
\label{fir}
\caption{Comparison between the total infrared luminosity inferred with our model from
the observed L-band emission,
assuming standard bolometric corrections (see text for details), and the luminosity measured by IRAS.}
\end{figure}

\rm

\subsubsection{Comparison with previous results}

%The model described above allows a quantitative estimate of the relative AGN/SB 
%contribution in ULIRGs.
Restricting our analysis on the complete sample made of the ULIRGs
listed in Table~6 (14 sources, 4 with resolved double nuclei)
we note that an AGN was found in 9 cases. In two cases the AGN is 
the dominant contribution to the bolometric luminosity, while in the other 7 cases the
starburst is dominant. 
In the following we compare these results with the classification provided by studies at other
wavelengths (Table~5):
\begin{itemize}
\item {\bf X-rays:} the classification is in all cases in agreement with the
results of  X-ray spectroscopy.
\item {\bf Mid-IR:} the mid-IR analysis of G98, based on coronal
lines and PAH emission features in ISO spectra, is in agreement with our conclusions
for the five starburst ULIRGs. Signatures of an AGN are clearly found
in two sources (MKN~231, MKN~273). One source (IRAS~05189-2524)
is not in the G98 sample. One source (IRAS~19254-7245) show marginal evidence 
of the presence of an AGN (its position in the diagnostic diagrams is ``half way''
the starburst and AGN regions, but a reanalysis of Charmandaris et al.~(2003) show
clear evidence of the AGN). The remaining four sources (IRAS~20551-4250, UGC~5101,
IRAS~23128-5919, NGC~6240) are classified as pure starbursts. 
\item {\bf Optical/Near-IR:} spectral evidence of the presence of an AGN
was found in four sources. All other objects have a starburst/LINER classification.
Among them, the objects with an L-band detected AGN are NGC~6240,
IRAS~20551-4250, IRAS~23128-5919, UGC~5101).
\end{itemize}
Summarizing, the above comparison shows that only L-band and hard X-ray spectroscopy
is effective in finding evidence of AGNs in starburst-dominated and/or heavily
absorbed ULIRGs, while the analysis at other wavelength fails in several cases.

This result can be easily explained comparing the fractions of {\it L-band}
luminosity due to AGN, $\alpha$, and the fraction of {\it bolometric} luminosity
due to AGN, $\alpha_{BOL}$ (Table~6): in ALL cases where an AGN is present,
its contribtuion to the L-band emission is dominant, even if this is not the case
for the bolometric emission in 7 out of 9 cases.
The same consideration applies to the hard X-ray band (E$>$2~keV), due to the much higher
fraction of bolometric luminosity emitted in the X-rays by AGNs (\~several percent) 
than by starbursts ($<10^{-4}$), but not to any other spectral region:\\
- at wavelengths between the hard X-rays and 2-5~$\mu$m the AGN emission is in
almost all cases completely absorbed by dust;\\
- at wavelengths higher than a few micron the AGN emission is diluted 
by the dominant starburst emission.

Concluding, the L-band and hard X-ray spectroscopy are the best spectral regions in
order to investigate the energy source in ULIRGs. As pointed out in
Section~6.1, the X-ray band analysis is limited to a handful sources with the current
instrumentation. Therefore, L-band spectroscopy is the most promising way
to extend our understanding of the AGN/starburst nature of ULIRGs in larger samples.
{We stress again that the diagnostics developed here has been suggested by the analysis 
of our high quality ISAAC spectra, but once calibrated does not require particularly
high signal-to-noise data, and is capable to solve cases which were not easily understandable
in previous analyses. A particularly striking example is the luminous infrared galaxy
NGC~4418 ($L_{IR}=10^{11.2}~L_\odot$), which show clear signatures of the presence of
an AGN in the mid-IR (\citealt{dud-win}, \citealt{spoon}), but not in the L-band, 
according to \citet{im4418}. However, when our diagnostic diagrams (Fig.~3a and 3b) 
are applied to its L-band spectrum, we find that it clearly lays in the AGN region.
Our model indicates an interpretation similar to that for IRAS~20551-4250, with a 
heavily buried AGN ($\tau\sim3-4$) and a starburst component responsible for $\sim$70\% of
the bolometric luminosity.}    

\subsection{Composition of the circumnuclear medium in ULIRGs}

We already noticed that the correction used to estimate the intrinsic
L-band luminosity in heavily obscured AGNs, based on the 
optical depth of the 3.4~$\mu$m absorption feature, leads to values
$\log R_L^{corr}>0$ for three out of the 4 sources with $\tau_{3.4}>0.3$. 
In particular,
we obtain $\log R_L^{corr}=2.7$ for IRAS~19254-7245, $\log R_L^{corr}=3.2$
for IRAS~08572+3915, $\log R_L^{corr}=2.3$ for UGC~5101.
These unphysically high values can be discussed considering their
relation (a) with the L-band continuum optical depth $\tau_L$
estimated with the model presented in the previous Section, and (b) 
with the gas absorption measured in the X-rays.

(a) The high optical depth of the 3.4~$\mu$m absorption features
in three objects implies an L-band extinction of several magnitudes,
assuming the standard absorption law of \citet{pend94}
($A_L\sim12\pm4 \tau_{3.4}$). Even considering the lowest allowed
value, i.e. $A_L=8 \tau_{3.4}$, we would still have an L-band to
total infrared ratio $\log R_L^{corr}>0$ for two sources (IRAS~19254-7245
and IRAS~08572+3915).

On the other hand, the extinction $\tau_L$ obtained from the spectral
decomposition presented in the previous Section is significantly
lower in the three objects with the highest $\tau_{3.4}$, and
compatible with an intrinsic L-band to bolometric ratio $\log R^{corr}_L\sim 0$. 
The comparison between the two estimates of the optical depth is
shown in Table~6.

The above results imply that 
the abundance of hydrocarbon dust grains, responsible for
the 3.4~$\mu$m absorption, is on average higher in these ULIRGs than
in the Galactic ISM. This is the case for all the three sources
with strong absorption at 3.4~$\mu$m ($\tau_{3.4}\sim0.8-0.9$).
We note that this is not necessarily a general feature in ULIRGs,
since selection effects could play a role: such high $\tau_{3.4}$
can  be observed only in objects where the dust grains responsible
of the 3.4~$\mu$m absorption are overabundant, because
otherwise the continuum extinction would be so high to prevent the AGN emission
from being observable.

(b) The column density measured in the X-rays are $N_H>10^{24}$~cm$^{-2}$ for
UGC~5101 and IRAS~19254-7245, based on the {\em XMM-Newton} observations 
(\citealt{im5101x} and \citealt{braito03}) and
$N_H>10^{25}$~cm$^{-2}$ for IRAS~08572+3915 from
a BeppoSAX observation (Risaliti et al.~2005, in prep.). Assuming a standard
Galactic extinction curve and dust-to-gas ratio (\citealt{savage}) we have
$A_L\sim 1.2~N_{H,23}$, where $N_{H,23}$ is the column density in
units of $10^{23}$~cm$^{-2}$. This would imply $A_L>12$ for
UGC~5101 and IRAS~19254-7245, and $A_L>120$ for IRAS~08572+3915.
Again, these values are not acceptable, because they would imply a
too high intrinsic continuum. We conclude that the dust-to-gas ratio
in heavily obscured ULIRGs is at least a factor of a few smaller than
in the Galactic ISM. This is a well known property of the absorbing
medium in Seyfert Galaxies with moderate absorption, i.e. $N_H<10^{23}$~cm$^{-2}$ 
(\citealt{macc82}, \citealt{maio}). 
Here we find that this is also the case for higher luminosity
and higher column density AGNs.

\section{Conclusions}

In this paper two works are presented: 
(1) L-band spectra obtained with ISAAC at VLT are analyzed
for a sample of 7 Ultraluminous Infrared Galaxies, and (2) a
detailed discussion on L-band diagnostics for ULIRGs is
performed, using a complete sample of local bright
ULIRGs, obtained merging our observations with the 
available L-band spectra of sources in the northern hemisphere.

1. The spectral analysis of the sample of 7 bright southern
ULIRGs show the power of ISAAC in revealing the nature
of these objects. Six sources are known to have double nuclei.
In four cases we obtained a spectrum for both nuclei. 
A clear signature of the presence of an AGN was found in three 
nuclei (IRAS~19254-7245S, IRAS~20551-4250, IRAS~23128-3915S).
In two of these three objects (IRAS~19254-7245A, IRAS~20551-4250) 
an absorption feature at 3.4~$\mu$m, due to hydrocarbon dust
absorption was detected. These two sources have also the steepest
continuum ($\Gamma>2$). Water ice absorption at 3.1~$\mu$m
was also detected in several objects. 
We confirm that 3.4~$\mu$m absorption is a typical
signature of an AGN, while 3.1~$\mu$m absorption
with moderate optical depths ($\tau_{3.1} < 0.5$) is found in
sources with no indication of an active nucleus.

The high quality of the spectra obtained with ISAAC demonstrates
that L-band spectroscopy can be a powerful tool to investigate
the nature of relatively faint ULIRGs. For example, most of
the 118~sources of the 1~Jy sample (\citealt{kim98}) are
probably bright enough for such a study. This is not true for
studies at other wavelengths, such as in the X-rays, which are in
principle extremely effective in disentangling AGNs and starbursts
but can only be performed with the 10/15 brightest
sources with the currently available instrumentation. 

2. We performed a complete analysis of the L-band spectral 
diagnostics on a sample consisting of our seven southern
ULIRGs, 7 northern ULIRGs selected with the same criterion
(60~$\mu$m flux density $S_{60}>5.2$~Jy) and 5 more sources
which are known to be either AGN-dominated or starburst
dominated, with a slightly lower bolometric luminosity. All these
sources have a reliable, independent starburst/AGN classification
based on observations at other wavelengths.
The main results are the following:
\begin{itemize}
\item In all known AGN, {\em at least} one L-band AGN indicator
is detected, while no AGN indicators are found in any object classified
as starburst. However, no single indicator provides a 100\% correct
classification.
\item When more indicators are combined, a 100\% correct starburst/AGN
classification is possible. We found that AGNs and starbursts are 
completely separated in two different bi-dimensional diagrams, the first obtained plotting
the 3.5-4~$\mu$m continuum slope, $\Gamma$, versus 
the 3.3~$\mu$m PAH emission feature equivalent width, $EW_{3.3}$; the second plotting
$EW_{3.3}$ versus the extinction-corrected L-band to bolometric emission ratio.
\item ULIRGs hosting AGNs can be clearly separated into two groups. The
first consists of objects 
where the AGN is not heavily absorbed in the L-band. The continuum is dominated
by the re-emission by hot dust of the AGN direct emission; 
the continuum slope (in a $\lambda-f_\lambda$
plane) is flat; the PAH emission feature is weak or absent; absorption features 
are absent. The second group is made by objects heavily absorbed in the L-band.
In these cases the emission is dominated by by the reprocessing by cold dust. The 
continuum is steep ($\Gamma\sim2$) due to reddening, and strong absorption features are present.
\item We use a simple model for the spectral deconvolution of the AGN and starburst components.
We show that it is possible to estimate the relative AGN and starburst contributions
to the bolometric luminosity through L-band spectral indicators.
\item The application of our model to a representative sample of ULIRGs in
the local Universe shows that AGNs are present in $\sim60$\% of ULIRGs. In
about half of these cases the AGN contribution to the bolometric luminosity is dominant.
Overall, the AGN contribution to the ULIRG luminosity in the local Universe
is $\sim30$\%, the remaining 70\% being due to starburst emission.
\item The comparison between gas absorption (from X-ray observations), continuum
extinction in the L-band, and the narrow absorption feature at 3.4~$\mu$m suggest that heavily 
obscured ULIRGs have a low dust-to-gas ratio and a different extinction curve
with respect to the Galactic ISM. In particular, hydrocarbon dust grains are overabundant
in three out of four cases with respect to the Galactic dust composition.
\end{itemize}

\section*{Acknowledgments}
We are grateful to Loredana Bassani for her contribution to the early
stage of this project, and to the anonymous referee for very useful comments.
This publication makes use of data products from the Two Micron All Sky Survey, 
which is a joint project of the University of Massachusetts and the Infrared Processing 
and Analysis Center/California Institute of Technology, funded by the National 
Aeronautics and Space Administration and the National Science Foundation.
Also, we made use of the NASA/IPAC Extragalactic Database (NED) which is operated 
by the Jet Propulsion Laboratory, California Institute of Technology, 
under contract with the National Aeronautics and Space Administration.

\appendix
\section{Analysis of the AGN/SB deconvolution model}

The relations plotted in Fig.~5, and expressed in Equations~3 and~4
can be inverted in order to work out the explicit dependence
of the two model parameters, $\alpha$ and $\tau_L$, on the
observed parameters, $EW$ and $\Gamma$:
\begin{equation}
\tau_L=\frac{EW_{SB}\Gamma-EW}{\beta(EW_{SB}-EW})
\end{equation}
\begin{equation}
\alpha=\frac{EW_{SB}-EW}{(EW_{SB}-EW)+EW*e^{-\tau_L}}
\end{equation}

where $\beta=1.75$ is the slope of the extinction curve in the L-band, 
$EW_{SB}$ is the intrinsic equivalent width of the 3.3~$\mu$m PAH feature
in a pure starburst spectrum.
 
From these two equations we obtain the errors on $\tau_L$ and $\alpha$:
\begin{equation}
\Delta\tau_L=\frac{EW}{EW_{SB}-EW} \frac{\vert\Gamma-1\vert\Delta(EW)}{\beta (EW_{SB}-EW)}
+\frac{\Delta(\Gamma)}{\beta}
\end{equation}
\begin{equation}
\Delta\alpha=\frac{\alpha^2 e^{-\tau_L}}{EW_{SB}-EW} EW\Delta(\tau_L)
+ \frac{EW_{SB}\Delta(EW)}{EW_{SB}-EW})
\end{equation} 
where $\Delta(EW)$ and $\Delta(\Gamma)$ are the errors on our measurements.

These equations have been used to estimate the errors in Table~6. Moreover,
they can be analyzed in order to quantitatively estimate in which regions of the
($EW, \Gamma$) parameter space we can measure the relative SB/AGN contribution
with good precision. 
In Fig.~A1 we plot the relative errors, $\Delta(\tau_L)/\tau_L$ and $\Delta(EW)/EW$
versus the measured $EW$, for different values of $\Gamma$, and assuming 
$\Delta(\tau_L)/\tau_L=0.1$ and $\Delta(EW)/EW$=0.2.

The regions with a bad determination of $\alpha$ (with errors higher than 50\%)
are those with high equivalent width of the 3.3~$\mu$m emission feature and
a steep spectrum, corresponding to the upper right 
region in Fig.~5 where the model lines get closer while approaching the pure starburst
point. This is also clear from Eq.~A4, where the value of $\Delta\alpha$
diverges for $EW\rightarrow EW_{SB}$.

\begin{figure}
\includegraphics[width=8cm]{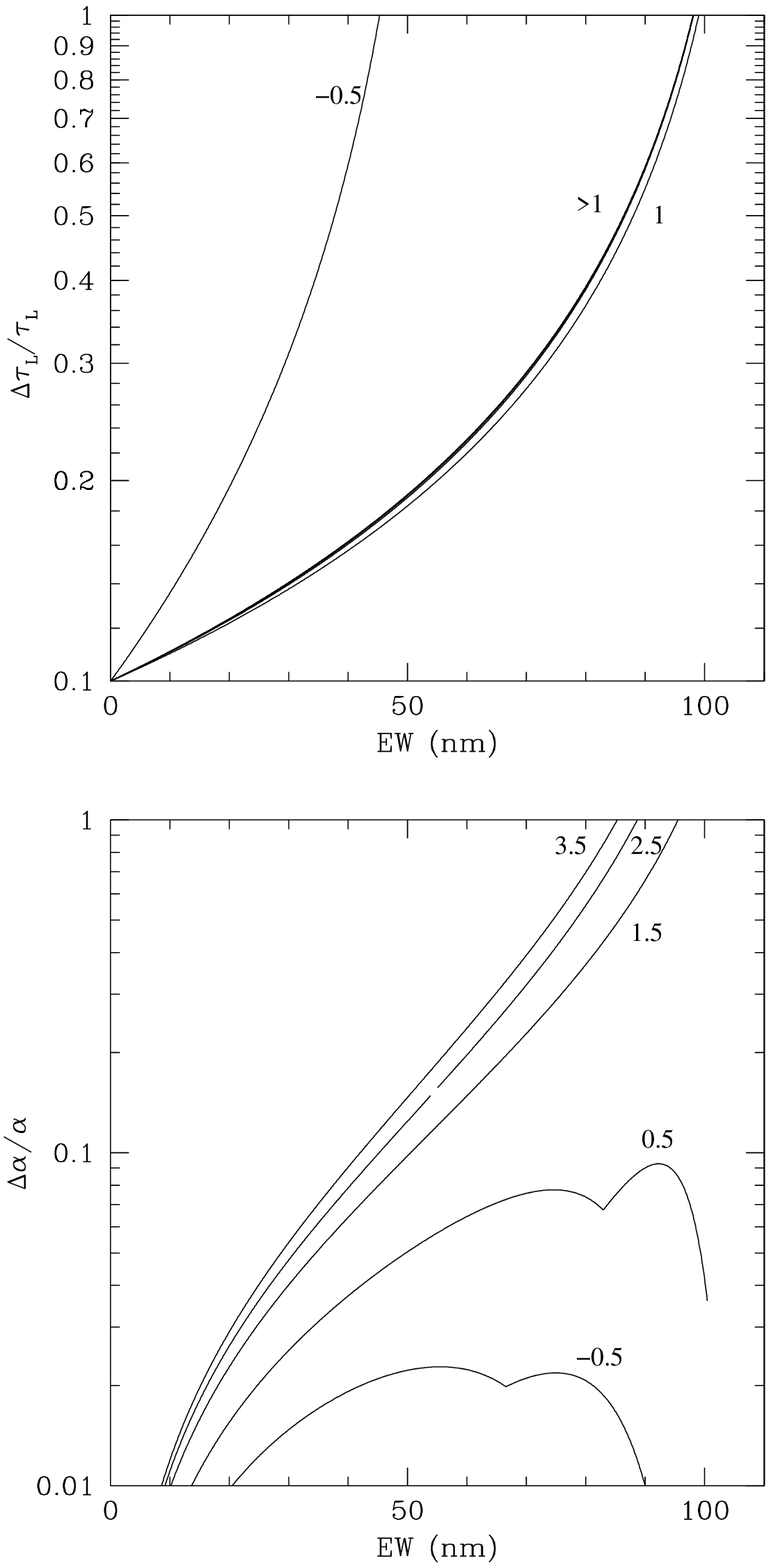}
\caption{Relative errors on $\tau_L$ (upper panel) and $\alpha$ (lower panel)
versus the observed equivalent width of the 3.3~$\mu$m emission feature, for different values
of the observed $\Gamma$. The value of $\Gamma$ is labeled in each line.} 
\end{figure}

\end{document}